\documentclass[11pt]{article}
\pdfoutput=1

\usepackage[letterpaper,
width=6.5in, 
left=1in,
top=1.3in,
bottom=1.15in,
headheight=19pt,
]{geometry}

\usepackage[utf8]{inputenc} % allow utf-8 input
\usepackage[T1]{fontenc}
\usepackage{url}            % simple URL typesetting
\usepackage{booktabs}       % professional-quality tables
\usepackage{cite}
\usepackage{graphicx}
\usepackage{floatflt}
\usepackage{amssymb}
\usepackage{amsmath}
\usepackage{mathtools}
\usepackage{xfrac}
\usepackage{xspace}
\usepackage{enumitem}
\usepackage[usenames,dvipsnames]{color}
\usepackage{soul}
\usepackage{tikz}
\usepackage[normalem]{ulem}
\usepackage{wrapfig}
\usepackage[hang,flushmargin, bottom]{footmisc}
\usepackage{algorithm, algorithmicx}
\usepackage{multirow}
\usepackage[square,sort,comma,numbers]{natbib}
\usepackage{units}

\usepackage{caption}
\DeclareCaptionFont{xipt}{\fontfamily{fvs}\fontsize{8}{10}\selectfont}
\usepackage[labelfont={xipt,bf},textfont={xipt},font={xipt}]{caption}

\DeclarePairedDelimiter\ceil{\lceil}{\rceil}
\DeclarePairedDelimiter\floor{\lfloor}{\rfloor}

\newfloat{messagedef}{H}{info}
\floatname{messagedef}{Message}
\newfloat{algorithmdef}{H}{info}
\floatname{algorithmdef}{Algorithm}

\setlist{nolistsep} 
\definecolor{darkgreen}{RGB}{0,192,0}
\definecolor{steelblue}{RGB}{182,221,232}
\definecolor{lightorange}{RGB}{252,213,180}
\definecolor{lightmagenta}{RGB}{204,192,218}
\definecolor{gray}{RGB}{128,128,128}

\newcommand{\Architecture}{Flow\xspace}
\newcommand{\SPoCK}{Specialized Proof of Confidential Knowledge\xspace}
\newcommand{\techterm}[1]{{\sffamily\selectfont{#1}}}

\graphicspath{{./}{./figures/}{./graphics/}}

\usepackage{ntheorem}
\theoremstyle{plain}
\theorembodyfont{\normalfont\sffamily\upshape\selectfont}
\newtheorem{theorem}{Theorem}
\newtheorem{corollary}{Corollary}

\newtheorem{definition}{Definition}

\usepackage{hyperref}       % hyperlinks
\hypersetup{
	colorlinks=true,       % false: boxed links; true: colored links
	linkcolor=[rgb]{0,0.4,0.5},          % color of internal links (change box color with linkbordercolor)
	citecolor=[rgb]{0,0.3,0.3},        % color of links to bibliography
	urlcolor=[rgb]{0,0.3,0}           % color of external links
}

\title{
	\Architecture: Separating Consensus and Compute\\[5pt]\Large 
	-- Execution Verification --
}
\date{} % Omitting the date when using \maketitle

\begin{document}
	\maketitle

\begin{center}
	Dr.\ Alexander Hentschel \\
	\footnotesize\texttt{alex.hentschel@dapperlabs.com} \\[10pt]
	\large
	\begin{tabular}{ccc}
		Maor Zamski & Dieter Shirley & Layne Lafrance \\[-3pt]
		\footnotesize\texttt{maor@dapperlabs.com} & \footnotesize\texttt{dete@dapperlabs.com} & \footnotesize\texttt{layne@dapperlabs.com} 
	\end{tabular}
\end{center}
\vspace{10pt}

\begin{abstract}
		\noindent
		Throughput limitations of existing blockchain architectures are well documented and are one of the most significant hurdles for their wide-spread adoption. 
		In our previous proof-of-concept work, we have shown that separating computation from consensus can provide a significant throughput increase
		without compromising security \cite{Bamboo:2019:SeparatingConsensusAndCompute}. 
		In our architecture, \techterm{Consensus Nodes} only define the transaction order but do not execute transactions.
		Instead, computing the block result is delegated to compute-optimized \techterm{Execution Nodes}, 
		and dedicated  \techterm{Verification Nodes} check the computation result.
		During normal operation, \techterm{Consensus Nodes} do not inspect the computation but oversee that participating nodes execute their tasks with due diligence and
		adjudicate potential result challenges.  
		While the architecture can significantly increase throughput, \techterm{Verification Nodes} still have to duplicate the computation fully.
		
		In this paper, we refine the architecture such that result verification is distributed and parallelized across many Verification Nodes.  
		The full architecture significantly increases throughput and delegates the computation work to the specialized \techterm{Execution Nodes}
		and the onus of checking it to a variety of less powerful \techterm{Verification Nodes}.
		We provide a full protocol specification of the verification process, including challenges to faulty computation results 
		and the resulting adjudication process. Furthermore, we formally prove liveness and safety of the system. 
\end{abstract}

\newpage
\tableofcontents

\newpage
\section{\Architecture Architecture}
\noindent
In most traditional blockchains, each full node must perform every task associated with running the system.
This process is akin to single-cycle microprocessors, where one instruction is executed per step.
In contrast, modern CPU design leverages pipelining to achieve higher throughput and scaling. 

Rather than asking every node to choose the transactions they will include in a block, compute the block’s output, 
come to a consensus on the output of those transactions with their peers, and finally sign the block, appending it onto the chain,
\Architecture adopts a pipelined architecture. In \Architecture, different tasks are assigned to specialized node roles:
\techterm{Collection}, \techterm{Consensus},  \techterm{Execution}, \techterm{Verification}, and \techterm{Observation}. 
This design allows high levels of participation in \techterm{Consensus} and \techterm{Verification} 
by individuals on home internet connections while leveraging large-scale datacenters to do most of the heavy lifting of \techterm{Execution}. 
Actors participating in \techterm{Consensus} and \techterm{Verification}  hold the other nodes accountable with crypto-economic incentives 
allowing \Architecture to gain massive throughput improvements without undermining the decentralization or safety of the network.
In a system with \techterm{Consensus} and \techterm{Execution} nodes only, 
\Architecture achieved a throughput increase by a \emph{factor} of 56 
compared to architectures where  consensus nodes also perform block computation \cite{Bamboo:2019:SeparatingConsensusAndCompute}. 
\medskip

\noindent
In any design where actors other than the consensus nodes perform tasks, 
correct task execution is not covered by the safety guarantees of consensus. 
Therefore, the protocol must include dedicated components to ensure secure system operation,
even in the presence of a moderate number of malicious participants. 
In our first white paper \cite{Bamboo:2019:SeparatingConsensusAndCompute}, we formally analyzed the security implications of
delegating tasks to other actors than \techterm{Consensus Nodes}. 
The central result of our research was that transaction execution can be transferred to one group of nodes (\techterm{Execution Nodes}), 
and result verification to an independent group (\techterm{Verification Nodes}). 
The protocol must include the following components to ensure safety:
\begin{itemize}
	\item The Verifiers must be able to appeal (formally: submit a challenge) to \techterm{Consensus Nodes} if they detect a protocol violation.
	\item \techterm{Consensus Nodes} must have the means to determine whether the challenger or the challenged is correct (formally: adjudicate the challenge). 
\end{itemize}
When such mechanisms are included, the pipelined architecture  is as secure as a blockchain where all tasks are executed by all consensus nodes
\cite{Bamboo:2019:SeparatingConsensusAndCompute}.
In the present paper, we refine the architecture such that result verification is distributed and parallelized across many Verification Nodes.  
Furthermore, we specify the details of the different challenges and adjudication protocols for execution verification. 

\subsubsection*{Core Architecture Principles}

\noindent
Above, we noted that nodes must be able to appeal to \techterm{Consensus Nodes} if they detect a protocol violation.
For an appeal system to provide security guarantees  that protect from Byzantine attacks, 
the system must have the following  attributes.
\begin{itemize}
	\item \textit{Detectable}:
	A single, honest actor in the network can detect deterministic faults, and prove the error to all other honest nodes by asking them to recreate part of the process that was executed incorrectly.
	\item \textit{Attributable}:
	All deterministic processes in \Architecture are assigned to nodes using a \techterm{verifiable random function (VRF)} \cite{Micali:1999:VRFs}.
	Any detected error can be attributed to the nodes responsible for that process.
	\item \textit{Punishable}:
	Every node participating in the \Architecture network must put up a stake, which is slashed
	in case the node is found to exhibit Byzantine behavior.
	Reliably punishing errors via slashing is possible because all errors in deterministic processes are detectable\footnote{%
		In \Architecture, nodes check protocol-compliant behavior of other nodes by re-executing their work. 
		In most cases, verification is computationally cheap, with the noticeable exception of computing all transactions in a block. 
		We describe in section \ref{sec:ExecutionFlow:Verification} how verifying the block computation is distributed and parallelized  
		such that each Verifier only has to perform a small fraction of the overall block computation. 
	}
	and attributable.
	\item \textit{Recoverable}:
	The \Architecture protocol contains specific elements for result verification and resolution of potential challenges.
	These elements serves to deter malicious actors from \emph{attempting} to induce errors that benefit them more than the slashing penalty,
	as the probability of their erroneous result being committed is negligible.
\end{itemize}

\subsubsection*{Assumptions}

\begin{itemize}
	\item 
	We solely focus on \techterm{Proof of Stake} blockchains, where all participants are known and each node is authenticatable through its signature.
	\item 
	Nodes commit to (and stake for) participating in the network for a specific time interval, which we refer to as an \techterm{Epoch}.
	Epochs are system-wide. 
	While nodes can participate over multiple Epochs, the end of an Epoch is a dedicated point in time for nodes to leave or join the system. 
	Epochs are considered to last for about a week.
	Technical details for determining the length of an Epoch 
	and a mechanism for Epoch changeover are left for future publications. 
	In this paper, we consider the system running only within one epoch.
\end{itemize}
\bigskip 

\noindent
Furthermore, we assume 
\begin{itemize}
	\item 
	The existence of a reliable
	source of randomness 
	that can be used for seeing  pseudo-random number generators. 
	We require the random seed to be unpredictable by any individual node until the seed itself is generated and published. 
	Possible solutions include \textsc{Dfinity}'s \techterm{Random Beacon} \cite{DFINITY:2018:Consensus}
	or proof-of-delay based systems \cite{buenz:2017:EthRandomness}.
	\item An aggregatable, non-interactive signature scheme, such as BLS signatures \cite{Boneh:2018:BLS:CompactMF}. 
    \item 
    Adequate compensation and slashing mechanics to incentivize nodes to comply with the protocol.
	\item 
	Partially synchronous network conditions with message traversal time bounded by $\Delta_t$. 
	Furthermore, we assume that local computation time is negligible compared to message traversal time. 
	\item 
	In numerous places throughout this paper, we refer to fractions of nodes. 
	This is a short-form of referring to a set of nodes which \emph{hold the respective fraction of stake}.
	Formally, let $\mathcal{N}^{(r)}$ be the set of all nodes with role $r$ and $s_\alpha$ the stake of node $\alpha \in \mathcal{N}^{(r)}$.
	A fraction of  \emph{at least} $\kappa$ nodes (with role $r$) refers to any subset
	\begin{align}
	\widetilde{\mathcal{N}} \subseteq \mathcal{N}^{(r)}  
	\textnormal{\quad such that }  
	\sum_{\tilde\alpha \in \widetilde{\mathcal{N}}} s_\alpha  \geq \kappa \hspace{-5pt} \sum_{\alpha \in \mathcal{N}^{(r)} } s_\alpha,
	\end{align}
	for $0 \leq \kappa \leq 1$.
	For example, stating that ``more than $\frac{2}{3}$ of Consensus Nodes have approved of a block'' implies that the approving nodes 
	hold more than  $\kappa = \frac{2}{3}$ of the Consensus Nodes' accumulated stake. 
\end{itemize}

\subsection{Roles}

\begin{figure}[b!]
	\centering
	\vspace{-25pt}
	\includegraphics[width=0.7\textwidth]{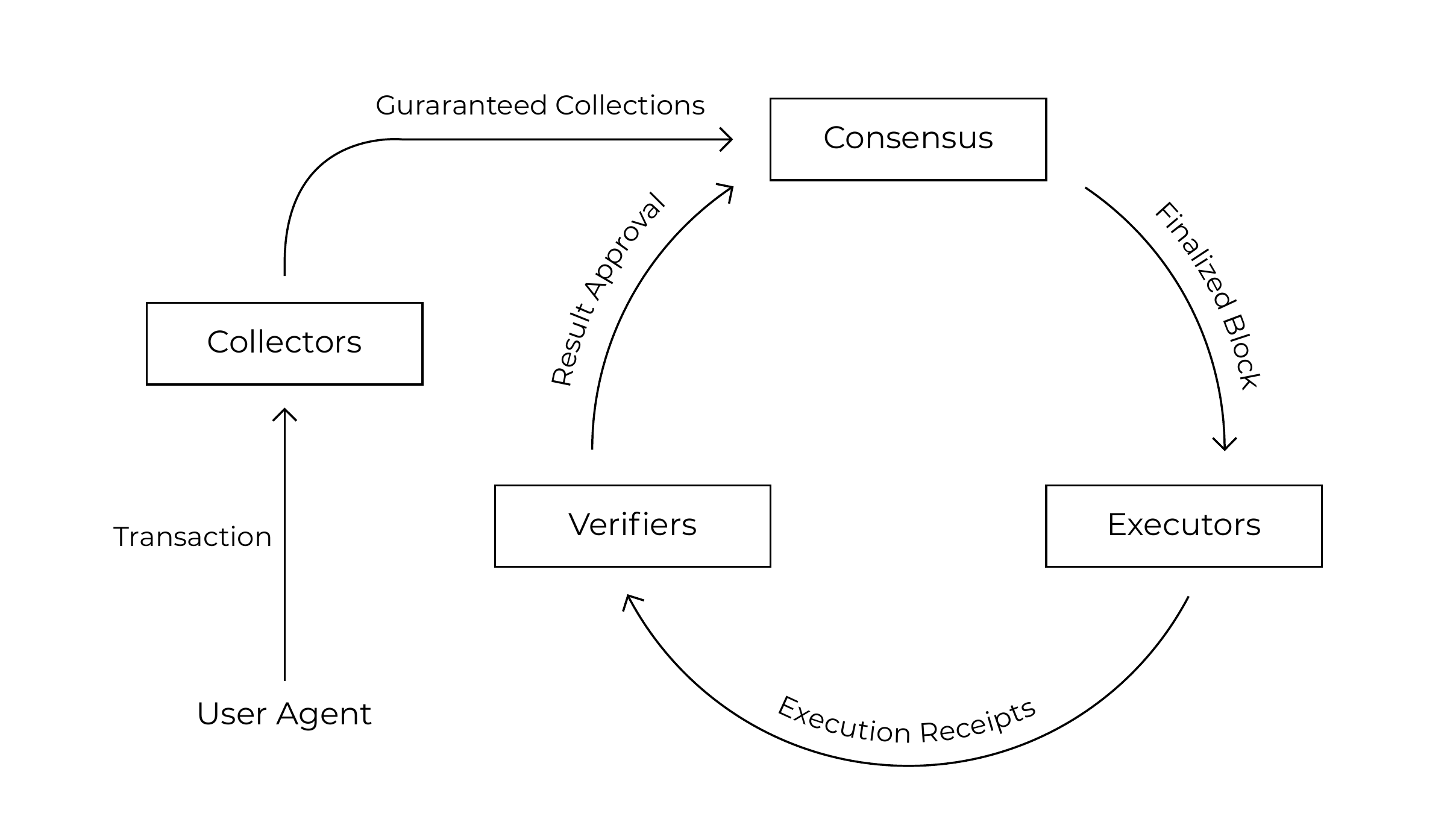}
	\vspace{-10pt}
	\caption{
		\textbf{Overview of the node roles and messages they exchange}.
		For simplicity, only the messages during normal operation are shown.
		Messages that are exchanged during the adjudication of slashing requests are omitted. 
	}
	\label{fig:Bamboo_Role_Overview}
\end{figure}

Roles are a service a node can provide to the network. These are: 
\techterm{Collector Role}, \techterm{Consensus Role}, \techterm{Execution Role}, \techterm{Verification Role}, and \techterm{Observer Role}. 
We refer to a network node that performs the respective role as \techterm{Collector Node}, \techterm{Consensus Node}, etc.
From an infrastructure perspective, the same hardware can host multiple roles in the network. However, the network treats individual roles 
as if they are independent nodes. Specifically, for each role a node stakes, unstakes, and is slashed independently. 
We furthermore assume that a node has it's own independent staking key for each of its roles, even if all roles are hosted on the same hardware. 
Figure \ref{fig:Bamboo_Role_Overview} illustrates the messages which nodes of the individual roles exchange during normal operation.

\subsubsection{Collector Role\label{sec:CoreArchitecture:Collectors}}

The central task of Collector Role is to receive transaction submissions from external clients and introduce them to the network. 
Staked nodes are compensated through transaction fees and all roles require a minimum stake to formally participate in that role. 
When receiving a transaction, a Collector Node checks that the transaction is well-formed. By signing the transaction, 
the node \techterm{guarantees} to store it until the transaction's result has been sealed.
(For details on block sealing, see section \ref{sec:ExecutionFlow:BlockSealing}.)
\medskip

\noindent
\textbf{\techterm{Clusters}:}
For the purpose of load-balancing, redundancy, and Byzantine resilience, Collection Nodes are partitioned into \techterm{clusters}. 
We require that Collection Nodes are staked equally. At the beginning of an Epoch, each Collection Node is assigned to exactly one cluster. 
Cluster assignment is randomized using the random beacon output.
\medskip

\noindent
\textbf{\techterm{Collections}:}
The central task of Collector Nodes is to collect well-formed transactions from external clients and to batch them into \techterm{collections}. 
When a Collector Node sees a well-formed transaction, it hashes the text of that transaction and signs the transaction to indicate two things: 
first, that it is well-formed, and second, that it will commit to storing the transaction text until the Execution Nodes have finished processing it. 
By  signing it, the Collector Node \techterm{guarantees} the transaction's storage and will subsequently be slashed (along with the rest of the cluster) if it doesn't produce the transaction. 

Collector nodes share all well-formed transactions they receive among their cluster and collaborate to form a joint collection of transactions. 
A cluster forms collections one at a time. Before a new collection is started, 
the current one is closed and send off to the Consensus Nodes for inclusion in a block. 
Further details on collections are provided in section \ref{sec:ExecutionFlow:CollectingTransactions}.

Collector Nodes in one cluster must agree on the transactions included in the current collection and at what point to close the collection. 
The determination of when to close a collection is based on a number of factors including token economics, which is out of scope for this paper. 
This distributed agreement  requires the nodes to run a consensus algorithm. 
Fortunately, the number of nodes in  a cluster and the transaction volume to processed by one cluster is moderate%
\footnote{
	For the mature system, we anticipate on the order of 20 to 50 nodes per cluster.
}.
Therefore, established  BFT  consensus algorithms, such as Tendermint \cite{Kwon2014TendermintC,  buchman2016tendermint, KwonAndBuchman:CosmosWhitepaper:2016},
SBFT \cite{Abraham_etal:2018:SBFT}, or Hot-Stuff \cite{HotStuff:2018}, are fully sufficient.
\medskip

\noindent
\textbf{Security implication:} 
As there are relatively few collectors in each cluster, a cluster may be compromised by malicious actors. 
The cluster could withhold the collection content that is referenced in a block but whose execution is still pending. 
The mitigation strategy (\hyperref[sec:AttackMitigation:MissingCollection]{\techterm{Missing Collection Challenge}}) for this attack is described in section \ref{sec:AttackMitigation:MissingCollection}.

\subsubsection{Consensus Role\label{sec:CoreArchitecture:Consensus}}

The Consensus Node's central tasks are the following:
\smallskip

\noindent
\textbf{\techterm{Block Formation}:}
Consensus Nodes form blocks from the collections. Essentially, Consensus Nodes maintain and extend the core \Architecture blockchain. 
In \Architecture, a block defines
the transactions as well as the other inputs (incl.\ the random seed) required  to execute the computation,
but not the resulting computational state \emph{after} block execution. 

An agreement to accept a proposed block needs to be reached by many nodes 
which requires a \techterm{Byzantine-Fault-Tolerant (BFT)} consensus algorithm
\cite{Wensley:1989:Fault-tolerant-computing, Reaching_Agreement_in_Presence_of_Faults:Pease:1980}.
While the specific design of the \Architecture's consensus system is still subject to active research, we restrict the design space to algorithms with the following guarantees. 
\begin{itemize}[listparindent=\parindent]
	\item 
	The algorithm is \emph{proven BFT}, 
	i.e.,\ it maintains consistency among honest participants, as long as less than one-third of Consensus Nodes are Byzantine. 
	\item 
	\emph{Safety} is guaranteed even in cases of network asynchrony. 
	Per definition of safety \cite{Lamport:1983:WeakByzantineGeneralsProblem, Howard:2018:DistributedConsensus}, 
	Consensus Nodes might declare a block finalized at different points in time, 
	depending on the amount of information they gathered from other Consensus Nodes. 
	Nevertheless, a safe consensus algorithm guarantees that all honest nodes eventually will declare the same block as finalized.
	\item 
	\emph{Liveness} is guaranteed in  partially synchronous network conditions \cite{DworkLynchStockmeyer:PartialSynchronyConsensus:1988}. 
	As the Fischer-Lynch-Paterson (FLP) theorem states, 
	a fault-tolerant, distributed consensus system cannot guarantee safety and liveness at the same time under 
	asynchronous network conditions \cite{FischerLynchPaterson:1985:FLP-Theorem}. 
	For \Architecture, we \emph{prioritize safety} over liveness in case of a network split. 
	Consistency of the world-state is more important than forward-progress of the network in extremely rare and adverse circumstances of a large-scale network split. 
	\newpage
	\item 
	A desired core feature of \Architecture's consensus algorithm is \emph{deterministic finality}.
	Once a block is finalized by \emph{any} Consensus Node, this commitment will never\footnote{
		Deterministic finality is guaranteed via BFT consensus, unless the system is under active attack of at least one-third of Byzantine actors.
	} be changed.
	Deterministic finality provides the significant advantage that dapp developers do not have to deal with the 
	effects of chain reorganizations. 
	\item \emph{Sustainability}: we do not consider proof-of-work consensus systems due to their exorbitant energy consumption.
	We focus on Proof of Stake systems only, where computational costs are reduced to necessary elements: primarily cryptography and book-keeping.		
\end{itemize}
\smallskip

\noindent
\textbf{\techterm{Block Sealing}:}
After a block has been computed and the resulting computational state verified, 
the  Consensus Nodes publish a \techterm{Block Seal} for the respective block. 
A Block Seal contains a commitment to the resulting computational state after block execution. 
Furthermore, it proves that the computation has been verified by a super-majority of Verifier Nodes. 

\begin{figure}[t!]
	\centering
	\includegraphics[width=\textwidth, trim=25 25 25 25, clip]{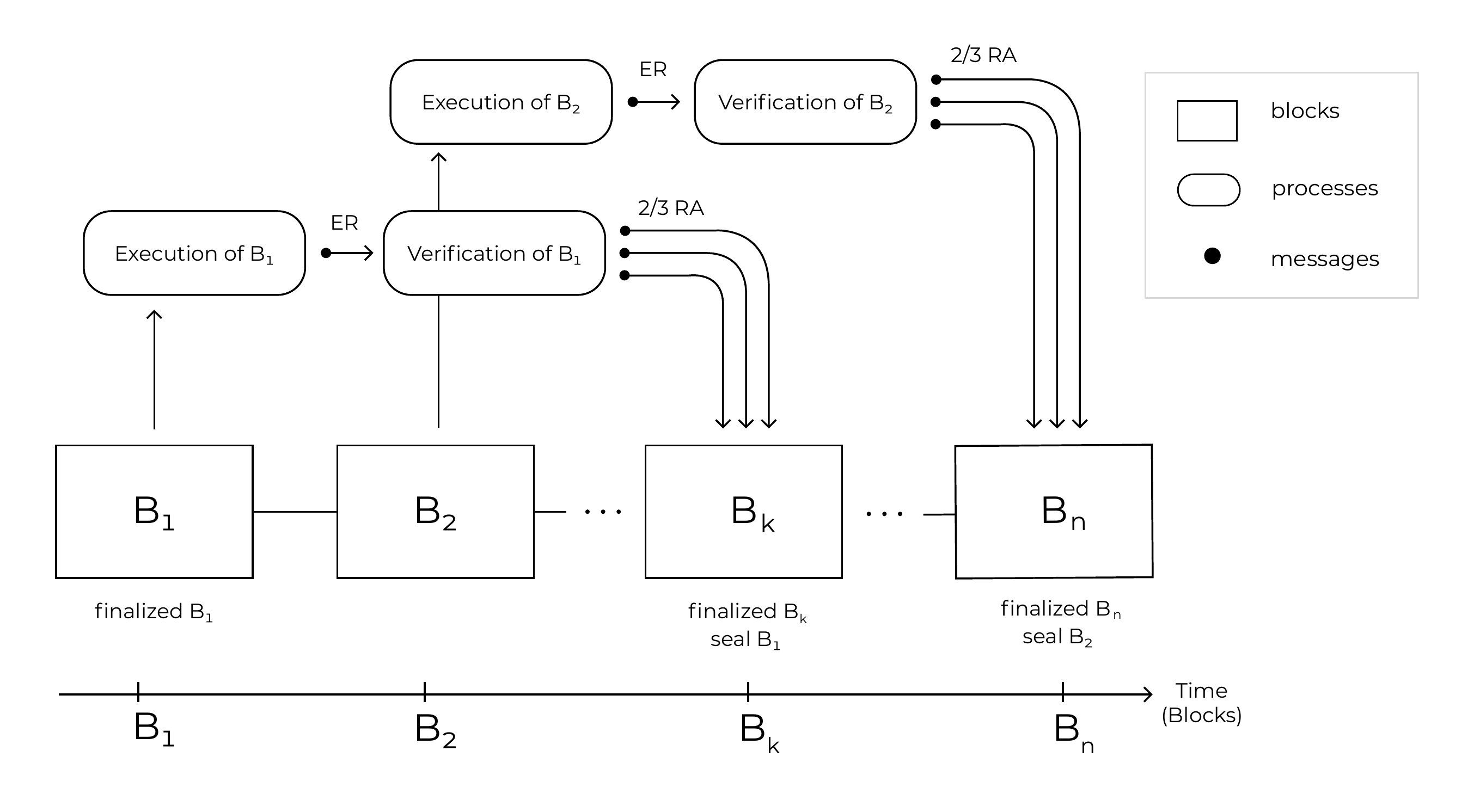}
	\caption{
		\textbf{Illustration of the placement of Block Seals within the \Architecture blockchain.} 
		After block $B_1$ is finalized by the Consensus Nodes, the Execution Nodes process their transactions
		and issue \techterm{Execution Receipts} (see section \ref{sec:ExecutionFlow:Execution}), 
		which are subsequently checked by the Verifier Nodes (see section \ref{sec:ExecutionFlow:Verification}).
		Provided that the checked parts of the computation result are found valid, a Verifier sends a  \techterm{Result Approval} 
		back to the Consensus Nodes. 
		As soon as conditions for sealing block $B_{1}$ are satisfied (see section \ref{sec:ExecutionFlow:BlockSealing}),
		the seal for the computation result of block $B_1$ is included in the next block $B_k$ that is generated by the Consensus Nodes.
	}
	\label{fig:BlockSealingIllustration}
\end{figure}

Consensus Nodes publish the block seals as part of the new blocks they finalize. 
As executing a block's transactions follows its finalization, the seal for the block's computational result 
cannot be included in the block itself. Instead, the seal is included in a later block. 
An illustration is shown in Figure \ref{fig:BlockSealingIllustration}.
\medskip

\noindent
\textbf{Tracking Staked Nodes}:
Consensus Nodes maintain a table of staked nodes including the 
nodes' current stakes and their public keys. 
It is important to note that staking balances are tracked solely by the Consensus Nodes 
and are \emph{not} part of the computational state.
Whenever a node's stake changes, Consensus Nodes publish this update as part of their next finalized block. 

During normal operations, staking and unstaking can only take place  at the switchover from one Epoch to the next.
However,  involuntary changes of stake through slashing can occur within an epoch and are accounted for 
by all honest nodes as soon as they process the block containing the respective stake update. 
\medskip

\noindent
\textbf{Slashing}:
Consensus Nodes adjudicate slashing challenges and adjust the staking balances accordingly.
\medskip

\noindent
\textbf{Security implication:} 
The consensus committee is the central authority in the system. 
The consensus committee itself must adjudicated challenges against committee members.
Safety and liveness are of this process are guaranteed through the BFT consensus algorithm. 
\Architecture uses a consensus algorithm with \emph{deterministic finality}. 
Therefore, dapp developers do not have to deal with the additional complexity of \techterm{chain reorganization}. 
Our results hold for \emph{any} BFT consensus algorithm with deterministic finality. 
\techterm{HotStuff} \cite{HotStuff:2018, HotStuff:2019:ACM} is the leading contender. 
However, we continue to assess other algorithms such as 
\techterm{Casper CBC} \cite{Zamfir:CasperCBC_Template:2017, Zamfir:CasperTFG:2017, Zamfir_et_al:MinimalCasperFamily:2018}
or \techterm{Fant\protect\^omette} \cite{Azouvi:2018:Fantomette}.

\subsubsection{Execution Role\label{sec:CoreArchitecture:Execution}}

Execution Nodes compute the outputs of all finalized blocks they are provided. 
They then ask the Collector Nodes for the collections containing transaction that are to be executed. 
With this data they execute the block and publish the resulting computational state as an \techterm{Execution Receipt}.
For verification, the computation is broken up into \techterm{chunks}. The Execution Nodes
publish additional information (see section \ref{sec:ExecutionFlow:Execution}) in the \techterm{Execution Receipt} about each chunk 
to enable Verification Nodes to check chunks independently and in parallel. 

The Computation Nodes are primarily responsible for \Architecture's improvements in scale and 
efficiency because only a very small number of these powerful compute resources are required 
to compute and store the canonical state.
\medskip

\noindent
\textbf{Security implication:} 
Malicious Execution Nodes could publish faulty \techterm{Execution Receipt}s. 
The protocol for detecting incorrect execution results is covered in section \ref{sec:ExecutionFlow:Verification}
and the adjudication process for the challenges in section \ref{sec:AttackMitigation:FaultyComputationResult}.

\subsubsection{Verification Role\label{sec:CoreArchitecture:Verification}}

Verification Nodes check the computation from the Execution Nodes.
While each node only checks a small number of chunks, all Verification Nodes together will check all chunks with overwhelming probability. 
For each chunk, a  Verification Node publishes a \techterm{Result Approval}, provided it agrees with the result. 
The \techterm{Execution Receipt} and the \techterm{Result Approval} are required for the block to be sealed. 
\medskip

\noindent
\textbf{Security implication:} 
Like most blockchains, \Architecture has to address the Verifier's Dilemma \cite{Luu:2015:VerifierDilemma}. 
In  a system where workers are producing results and Verifiers are confirming result correctness, 
there is an incentive for Verifiers to approve results without expending the work of checking. 
This conflict of incentives is at the heart of the Verifier's Dilemma.
It persists even the worker and Verifiers are \emph{not} colluding, so additional Verifiers do not help.
For \Architecture, we developed \hyperref[sec:Z]{\techterm{Specialized Proofs of Confidential Knowledge}}  (section \ref{sec:Z}) to overcome the Verifier's Dilemma
(section \ref{sec:ExecutionFlow:Verification:VerificationByRe-Computation} for more details).

\subsubsection{Observer Role\label{sec:CoreArchitecture:Observer}}
Observer Nodes relay data to protocol-external entities that are not participating directly in the protocol themselves.

\subsection{Locality of Information}

\begin{itemize}
	\item Correctness of information is cryptographically verifiable using on-chain information 
	\item \Architecture blockchain is, from a data perspective, \emph{not} self-contained.
				For example, cryptographic hashes of the computational state are included in blocks, but the state itself is not. 
	           This implies that anyone can verify the integrity of any subset of the computational state 
	           using the hashes in the \Architecture blockchain 
	           (and merkle proofs which must be provided alongside the actual data).
	           However, it is impossible to extract the state itself from the core blockchain. 
	\item Computational state is local to Execution Nodes
	\item Reference to the information holder is guaranteed by the holder's signature 
\end{itemize}

\subsection{Computational State vs.\ Network Infrastructure State\label{sec:Computational_vs_Network-State}}

An important conceptual difference in \Architecture is handling information that pertains to the computational state vs.\ that of the network infrastructure itself.  
While the computational state is held by the Execution Nodes, the network's infrastructure state is maintained by the Consensus Nodes. 

To illustrate the difference, consider the situation where the nodes in the network do not change (nodes never leave, join, or change  stake).
However, the transactions executed by the system will modify register values, deploy smart contracts, etc. In this setting, 
only the  computational state changes.  Its integrity is protected by the verification process (see section \ref{sec:ExecutionFlow:Verification} for details).

In contrast, let us now consider a situation where no transactions are submitted to the system. Blocks are still produced but contain no transactions. 
In this case, the system's  computational state  remains constant. However, when nodes leave or join, the state of the network infrastructure changes. 
The integrity of the network state is protected by the consensus protocol. To modify it, more than $\nicefrac{2}{3}$ of Consensus Nodes must approve it. 

\subsubsection{Staking and Slashing}

The network state itself contains primarily a list of all staked nodes which contains the node's staking amount and its public staking key. 
Updates to the network state are relevant for all nodes in the network. Hence, Consensus Nodes publish updates directly as part of the blocks they produce. 
Furthermore, slashing challenges are directly submitted to Consensus Nodes for adjudication.
(For example, section \ref{sec:AttackMitigation:FaultyComputationResult} provides a detailed protocol for challenging execution results.)
As Consensus Nodes maintain the network state, including staking balances,
they can directly slash the stake of misbehaving nodes, without relying on Execution Nodes to update balances.

\newpage
\section{General Techniques}
In this section, we describe methods and techniques used across different node roles.

\subsection[\SPoCK]{\SPoCK (\techterm{SPoCK})\label{sec:Z}}

A \techterm{SPoCK} allows any number of provers to demonstrate 
that they have the same confidential knowledge (secret $\zeta$).
The cryptographic proof does not leak  information about the secret. 
Each prover's \techterm{SPoCK} is specialized to them, and can not  be copied or forged without possession of the secret. 

The \techterm{SPoCK}  protocol is used in \Architecture to circumvent the Verifier's Dilemma \cite{Luu:2015:VerifierDilemma} (section \ref{sec:ExecutionFlow:Verification:VerificationByRe-Computation}).
The protocol prevents Execution or Verification Nodes from copying \techterm{Execution Results} or \techterm{Result Approvals} 
from each other.
Thereby, actors cannot be compensated for work they didn't complete in time.
In \Architecture, the secret $\zeta$ is derived from the \techterm{execution trace} of the low-level execution environment (e.g., the virtual machine). 
Executing the entire computation is the cheapest way to create the \techterm{execution trace}, even when the final output of computation is known.
\medskip

\noindent
Formally, the \techterm{SPoCK}  protocol provides two central guarantees:
\begin{enumerate}
	\item An arbitrary number of parties can prove that they have knowledge of a shared secret $\zeta$ without revealing the secret itself. 
	\item The proofs can be \emph{full revealed}, in an arbitrary order, without allowing any additional party to pretend knowledge of $\zeta$.
\end{enumerate}	
\smallskip

\noindent
The \techterm{SPoCK}  protocol  works as follows.
\begin{itemize}
	\item 
	Consider a normal blockchain that has a transition function $t()$ where $S' = t(B, S)$, 
	with $B$ being a block of transactions that modify the world state, $S$ as the state before processing the block, and $S'$ as the state after. 
	We create a new function $\tilde{t}()$ that works the same way, but has an additional secret output $\zeta$, 
	such that $(S', \zeta) = \tilde{t}(B, S)$ for the same $S$, $S'$, and $B$ as $t()$. 
	The additional output, $\zeta$, is a value \emph{deterministically} derived from performing the computation, 
	like a hash of the CPU registers at each execution step, which can't be derived any more cheaply than by re-executing the entire computation. 
	We can assume the set of possible values for $\zeta$ is very large.
	\item 
	An Execution Node, Alice, publishes a signed attestation to $S'$ (a merkle root of some sort), 
	and responds to queries about values in $S'$ with merkle proofs. Additionally it publishes a \texttt{SPoCK}  derived from $\zeta$.
	\item 
	A Verifier Node, Bob verifies that $S'$ is an accurate application of $t(B, S)$, and also publishes its own \texttt{SPoCK}  of $\zeta$.
	\item 
	An observer can confirm that both \texttt{SPoCK}s are derived from the same $\zeta$,
	and assume that Bob actually verified the output with high probability\footnote{%
		We assume that honest nodes will not accept unproved values for $\zeta$,
	    because they would be slashed if the $\zeta$-value was incorrect.
		Therefore, the observer can assume that statistically more than $\frac{2}{3}$ of the \techterm{SPoCK}s
		have been obtained by truthful re-computation of the respective  chunks.
	}
	and didn't just ``rubber stamp'' the result.
	This doesn't provide any protection in the case where Alice and Bob are actively colluding, 
	but it does prevent a lazy node from ``confirming'' a result without actually knowing that it is correct.
	\item 
	A \texttt{SPoCK}  is created as follows: 	
	\begin{itemize}
		\item 
		Use $\zeta$ (or a cryptographic hash of $\zeta$) as the seed for a deterministic key generation process, 
		generating a public/private key pair (\texttt{pk}, \texttt{sk}). 
		\item 
		Use the private key \texttt{sk} to sign your public identity (such as a node \texttt{ID}), 
		and publish the signature along with the deterministically generated public key \texttt{pk}: 
				 \begin{align}
				 		\textnormal{\texttt{SPoCK} }\ Z = (\texttt{pk}, \texttt{sign}_\texttt{sk}(\texttt{ID}))
				 \end{align}
	\end{itemize}
	All observers can verify the signatures to see that both Alice and Bob must have access to the private key \texttt{sk}, 
	but the signatures themselves don't allow recovery of those private keys. 
	Alice and Bob must both have knowledge of the same underlying secret $\zeta$ used to generate the private key.
\end{itemize}
In order t seal a block, several \techterm{SPoCK}s have to be validated for each chunk. 
Benchmarks of our early BLS implementation indicate that all proofs for sealing a block can be verified on a single CPU core in the  order of a second. 
Parallelization across multiple CPU cores is straight forward. 
Verification time can be further reduced by utilizing vectorized CPU instructions such as \techterm{AVX-512} or a cryptographic coprocessor.

\newpage
\section{Transaction Processing Flow}
\begin{figure}[h!]
	\centering
	\includegraphics[width=0.93\textwidth]{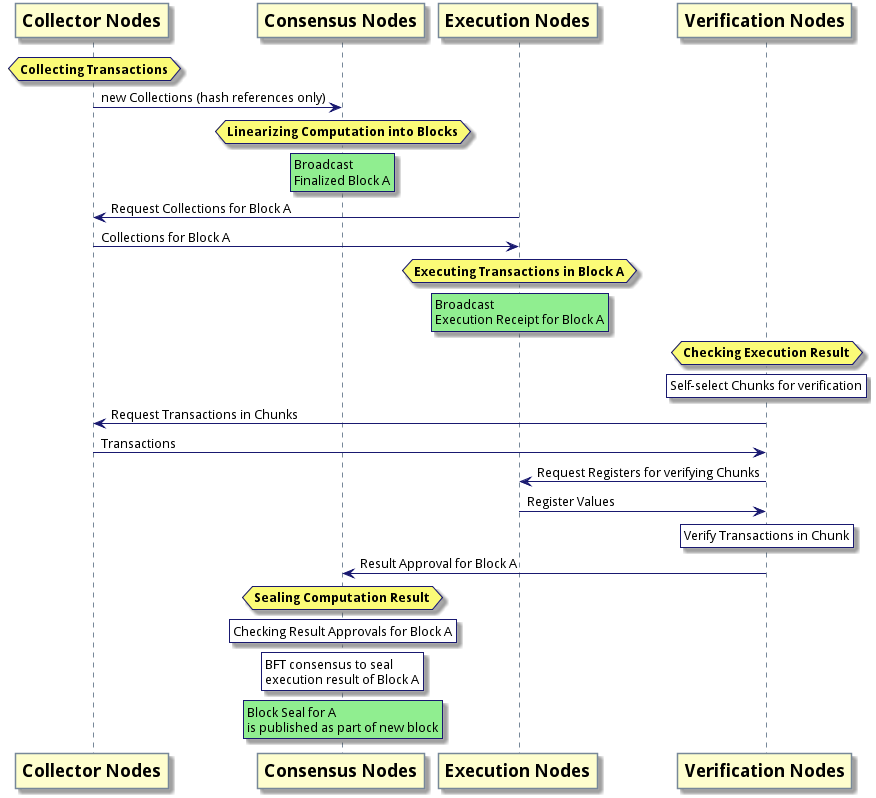}
	\caption{
		\textbf{Transaction Processing Flow}.
		The hexagon-shaped boxes indicate the start of the  individual stages, 
		which are discussed in sections \ref{sec:ExecutionFlow:CollectingTransactions} -- \ref{sec:ExecutionFlow:BlockSealing}.
		Arrows show message exchange between nodes with specific roles.
		Green boxes represent broadcast operations where the content is relayed to all staked nodes in the network (independent of their respective roles). 
		White boxes are operations the nodes execute locally.		
	}
	\label{fig:Post-Computation_Details}
\end{figure}

In this section, we present a formal definition of the transaction processing flow. 
Figure \ref{fig:Post-Computation_Details} provides a high-level overview of the individual stages 
which we discuss in detail in sections \ref{sec:ExecutionFlow:CollectingTransactions} -- \ref{sec:ExecutionFlow:BlockSealing}.
For conciseness and clarity, we focus on the core steps and omit protocol optimizations for conservation of bandwidth, run-time, etc. 
In section \ref{sec:ExecutionFlow:CorrectnessProof}, we formulate Theorem \ref{theorem:ExecutionFlow:CorrectnessProof} 
which proves that correctness of a sealed computation result is probabilistically guaranteed even in the presence of Byzantine actors.

To formally define the messages that nodes exchange, we use Protobuf-inspired\footnote{%
	Specifically, we use Protobuf with the exception of omitting the field numbers for the sake of conciseness.
}
pseudo-code  \cite{Google:Protobuf}.

\newpage
\subsection{Collector Role: Batching Transactions into Collections\label{sec:ExecutionFlow:CollectingTransactions}}

As described in section \ref{sec:CoreArchitecture:Collectors}, there are several clusters of collectors, where each cluster maintains one collection at a time. 
A cluster may decide to close its collection as long as it is non-empty. 
As part of closing the collection, it is signed by the cluster's collector nodes to indicate their agreement 
with the validity of its content and their commitment to storing it, until the block is sealed. 

\begin{wrapfigure}{R}{0.45\textwidth}
	\vspace{-15pt}
	\begin{minipage}{0.45\textwidth}
		\begin{algorithm}[H]
			\renewcommand{\thealgorithm}{}
			\floatname{algorithm}{}
			\begin{algorithmic}[1]
				\State \textbf{message} GuaranteedCollection $\{$
				\State \quad \textbf{bytes}  collectionHash;
				\State \quad \textbf{uint32} clusterIndex;
				\State \quad \textbf{Signature} aggregatedCollectorSigs; \label{msg:Collection:aggregatedCollectorSigs}
				\State $\}$
			\end{algorithmic}
		\end{algorithm}
	\end{minipage}
	\captionof{messagedef}{
		Collector Nodes send a \texttt{GuaranteedCollection} message  to Consensus Nodes to inform them about a guaranteed Collection. 
		For a collection to be guaranteed, more than $\frac{2}{3}$ of the collectors in the cluster (\texttt{clusterIndex}) must have signed it. 
		Instead of storing the signatures individually, \texttt{aggregatedCollectorSigs} is an aggregated signature.  
	}
	\label{msg:Collection}
	\vspace{-10pt}
\end{wrapfigure}

$~$\vspace{-15pt}
\begin{definition}
	A \textbf{Guaranteed Collection} is a list of transactions that is signed by more than 2/3rds of collectors in its cluster.
	By signing a collection, a Collector Node attests:
	\begin{itemize}
		\item that all transaction in the collection are well-formed;
		\item the collection contains no duplicated transactions;
		\item to storing the collection including the full texts of all contained transactions.
	\end{itemize}
\end{definition}

\noindent
A guaranteed collection is considered immutable. 
Once a collection is guaranteed by the collectors in its cluster, its reference is submitted to the Consensus Nodes for inclusion in a block
(see Message \ref{msg:Collection}).

\subsection{Consensus Role: Linearizing Computation into Blocks\label{sec:ExecutionFlow:LinearizingComputation}}

Consensus nodes receive the \texttt{GuaranteedCollection}s  from Collector clusters and include them in blocks
through a BFT consensus algorithm outlined in section  \ref{sec:CoreArchitecture:Consensus}.

\begin{figure}[b!]
\begin{center}
	\vspace{-20pt}
	\begin{minipage}{0.7\textwidth}
		\begin{algorithm}[H]
			\renewcommand{\thealgorithm}{}
			\floatname{algorithm}{}
			\begin{algorithmic}[1]
				\State \textbf{message} Block $\{$
				\State \quad \textbf{uint64} height;
				\State \quad \textbf{bytes} previousBlockHash;
				\State \quad \textbf{bytes} entropy;
				\State \quad \textbf{repeated GuaranteedCollection} guaranteedCollections;
				\State \quad \textbf{repeated BlockSeal} blockSeals; 
				\State \quad \textbf{SlashingChallenges} slashingChallenges; 
				\State \quad \textbf{NetworkStateUpdates} networkStateUpdates; 
				\State \quad  \textbf{Signature} aggregatedConsensusSigs;
				\State $\}$
			\end{algorithmic}
		\end{algorithm}
	\end{minipage}
	\vspace{-5pt}
	\captionof{messagedef}{
		Consensus Nodes broadcast \texttt{Block}s to the entire network.
		Their Block is valid if and only if more than $\frac{2}{3}$ of Consensus Nodes have signed. 
		Instead of storing the signatures individually, we use store an aggregated signature in \texttt{signedCollectionHashes}.  			
	}
	\label{msg:Block}
	\vspace{-15pt}
\end{center}
\end{figure}

The structure of a finalized block is given in Message \ref{msg:Block}. 
The universal fields for forming the core blockchain are \texttt{Block.height}, \texttt{Block.previousBlockHash}. 
Furthermore, the block contains a \texttt{Block.entropy} as a source of entropy, 
which is generated by the Consensus Nodes' Random Beacon.
It will be used by nodes that process the block to seed multiple random number generators
according to a predefined publicly-known protocol. 
In \Architecture, a reliable and verifiable source of randomness is essential for the system's Byzantine resilience. 

The filed \texttt{Block.guaranteedCollections} specifies the new transactions that are to be executed next. 
During normal operations, Consensus Nodes only require the information provided by the 
\hyperref[msg:Collection]{\texttt{GuaranteedCollection}} message.
In particular, they work only with the collection hash (\texttt{GuaranteedCollection.collectionHash}) but 
do not need to inspect the collection's transactions
unless an execution result is being challenged.

The \texttt{blockSeals} pertain to previous blocks in the chain whose execution results have just been sealed.  
A Block Seal contains a commitment to the resulting computational state after block execution 
and proves  the computation has been verified by a super-majority of Verifier nodes. 
Section \ref{sec:ExecutionFlow:BlockSealing} describes in detail how Consensus Nodes seal blocks. 

The fields \texttt{Block.slashingChallenges} and  \texttt{Block.networkStateUpdates} are listed for completeness only. 
For brevity, we do not formally specify the embedded messages \texttt{SlashingChallenges} and \texttt{NetworkStateUpdates}.
\texttt{Block.slashingChallenges} will list any slashing challenges that have been submitted to the Consensus Nodes
(but not necessarily adjudicated yet by the Consensus Nodes). Slashing challenges can be submitted by any staked node. 
Some challenges, 
such as the \techterm{Faulty Computation Challenge} discussed in section \ref{sec:AttackMitigation:FaultyComputationResult}), 
require the challenged node to respond and supply supplemental information.
The filed \texttt{networkStateUpdates} will contain any updates network infrastructure state 
(see section \ref{sec:Computational_vs_Network-State} for details). 
Most significantly, staking changes due to slashing, staking and unstaking requests are recorded in this field. 
By including \texttt{slashingChallenges}  and \texttt{networkStateUpdates} in a Block, their occurrences are persisted in the chain.
As all staked nodes must follow recently finalized blocks to fulfill their respective roles, journaling this information in the block also 
serves as a way to publicly broadcast it.

\subsection{Execution Role: Computing Transactions in Block\label{sec:ExecutionFlow:Execution}}

When the Execution Nodes receive a finalized Block, as shown in Message \ref{msg:Block}, they cache it for execution. 
An Execution Node can compute a block once it has the following information. 
\begin{itemize}
	\item 
	The \textit{execution result from the previous block} must be available, as it serves as the starting state for the next block. 
	In most cases, an Execution Node has also computed the prior block and the resulting output state is known.
	Alternatively, an Execution Node might request the resulting state from the previous block from a different Execution Node. 
	\item  
	The Execution Node must fetch \textit{all collections' text referenced in the block} from the Collector Nodes. 
\end{itemize}

\subsubsection{Chunking for  Parallelized  Verification\label{sec:ExecutionFlow:Execution:Chunking}}

\noindent
After completing a block's computation, an Execution Node broadcasts an \techterm{Execution Receipt} to the Consensus and Verification Nodes. 
We will discuss the details of the  \techterm{Execution Receipt} in more detail in section \ref{sec:ExecutionFlow:Execution:ExecutionReceipt}. 
At its core, the  \techterm{Execution Receipt} is the Execution Node's  commitment to its final result. 
Furthermore, it includes interim results that make it possible to check the parts of the computation in parallel without re-computing the entire block.
The execution result's correctness must be verified in a dedicated step to ensure an agreed-upon computational state. 
At the execution phase, the block computation is done by compute-optimised nodes.
We therefore assume that a block re-computation by any other node, as part of the independent verification process, will always be slower.
To address the bottleneck concern, we take an approach akin to split parallel verification \cite{Haakan:2002:ProbabilisticVerification}. 
We define the \techterm{Execution Receipt} to be composed of separate \techterm{chunks}, each constructed to be independently verifiable. 
A verification process (section \ref{sec:ExecutionFlow:Verification}) is defined on a subset of these chunks.

By breaking the block computation into separate chunks, 
we can distribute and parallelize the execution verification across many nodes.  
Let us consider a block containing $\Xi$ chunks. Furthermore, let $\eta$ be the fraction of chunks in a block that each Verification Node checks. 
The parameter-value for $\eta$ is protocol-generated  (see section \ref{sec:ExecutionFlow:CorrectnessProof} for details).
\begin{align}
	\textnormal{Number of chunks each Verification Node checks} = \ceil*{ \eta \ \Xi},
\end{align}
for $\ceil*{ \cdot}$ the ceiling function. 
A mature  system with  $\mathcal{V}$ Verification Nodes would re-compute in total $\mathcal{V} \ceil*{ \eta\   \Xi}$ chunks
if all Verification Nodes were to fully participate. Hence, on average, each chunk is executed 
$\frac{\mathcal{V} \ceil*{ \eta\   \Xi}}{\Xi}$ times. 
For example, a mature system with $\mathcal{V} = 1000$ Verification Nodes could break up a large block into $\Xi=1000$ chunks.
With $\eta = 4.2\%$, each Verification Node would check $\ceil*{ \eta \ \Xi} = 42$ chunks and each chunk would be verified by 
$42 = \frac{\mathcal{V} \ceil*{ \eta\   \Xi}}{\Xi}$ different Verification Nodes (on average).
The redundancy factor of $42$ makes the system resilient against  Byzantine actors (see section \ref{sec:ExecutionFlow:CompLoadOnVerifier}).

It is important that there is not single chunk that has a significant increased computation consumption compared to others in the block. 
If chunks had vastly different execution time,  Verifiers assigned to check the long-running chunks would likely not be finished before the block is 
sealed (see section \ref{sec:ExecutionFlow:BlockSealing}).
Hence, Execution Nodes could attack the network by targeting the long-running chunk to introduce a computational inconsistency that is left unverified. 
This weakness is mitigated by enforcing chunks with similar computation consumption.
Specifically, we introduce the following system-wide parameters.
\begin{align}
	\Gamma_{_{\hspace{-1pt}\textrm{Tx}}} : & \ \textnormal{ upper limit for the computation consumption of one transaction} \\
	\Gamma_{_{\hspace{-1pt}\textrm{chunk}}} : & \ \textnormal{ upper limit for the computation consumption of all transactions in one chunk}\\
	\textnormal{with } & \Gamma_{_{\hspace{-1pt}\textrm{Tx}}} \ll \Gamma_{_{\hspace{-1pt}\textrm{chunk}}} 
\end{align}
Here, we assume the existence of a measure of computation consumption similar to \techterm{gas} in Ethereum.

\begin{figure}[t!]
	\raggedleft
	\vspace{-10pt}
	\begin{algorithm}[H]
		\caption{Chunking}
		\textbf{Input:} 
		\begin{itemize}
			\item[$\mathcal{T}$:] List; element $\mathcal{T}[i]$ is  computation consumption of transaction with index $i$ in current block
		\end{itemize}
		\textbf{Output:}
		\begin{itemize}
			\item[$\mathcal{C}$:] 
			List; element $(k , c) \equiv \mathcal{C}[i]$ represents chunk with index $i$ in current block; 
			$k$ is index of first transaction in chunk; $c$ is chunk's computation consumption
		\end{itemize}			
		\begin{algorithmic}[1]
			\State \textbf{Chunking}($\mathcal{T}$) 
			\State \quad $\mathcal{C} \leftarrow [~]$ \Comment{initialize $\mathcal{C} $ as empty list}
			\State \quad $c\leftarrow 0$			\Comment{computation consumption of current chunk}	
			\State \quad $k\leftarrow 0$			\Comment{start index of current chunk}	
			\State \quad \textbf{for} $i = 0, 1, \ldots ,\texttt{length}(\mathcal{T})-1$:
			\State \quad \quad \textbf{if} $c + \mathcal{T}[i] > \Gamma_{_{\hspace{-1pt}\textrm{chunk}}}$: \label{alg:Chunking:add_decission} \Comment{adding transaction with index $i$ would overflow chunk}	
			\State \quad \quad \quad $e \leftarrow (k, c)$ \Comment{complete current chunk without transaction $i$}	
			\State \quad \quad \quad $\mathcal{C}$\texttt{.append}$(e)$
			\State \quad \quad \quad $k \leftarrow i$ \Comment{start next chunk at transaction $i$}	
			\State \quad \quad \quad $c \leftarrow \mathcal{T}[i]$
			\State \quad \quad \textbf{else}: \Comment{current chunk computation consumption of current chunk}	
			\State \quad \quad \quad $c \leftarrow c + \mathcal{T}[i]$
			\State \quad $e \leftarrow (k, c)$ \Comment{complete last chunk}	
			\State \quad $\mathcal{C}$\texttt{.append}$(e)$			
			\State \quad \textbf{return} $\mathcal{C}$
		\end{algorithmic}
	\end{algorithm}
	\vspace{-15pt}
	\captionof{algorithmdef}{
		separates transactions in a block into chunks. The algorithm presumes that computation consumption 
		$ \mathcal{T}[i] \leq \Gamma_{_{\hspace{-1pt}\textrm{Tx}}}$ for all $i$. 
	}
	\label{alg:Chunking}
\end{figure}

Since there is a hard limit of computation consumption both on a transaction as well as a chunk, 
with the chunk's being significantly higher than the transaction's, 
the simple \techterm{greedy Algorithm} \ref{alg:Chunking} will achieve the goal of similar computational consumption. 
Formally, let $\Gamma_{_{\hspace{-1pt}\textrm{chunk}}} = n \cdot \Gamma_{_{\hspace{-1pt}\textrm{Tx}}}$, for $n \ge 1$. 
Then, all chunks, except for the last one, will have a computation consumption $c$ with the following properties.
\begin{itemize}
	\item  
	$(n-1) \Gamma_{_{\hspace{-1pt}\textrm{Tx}}} < c$. 
	If this was not the case,  i.e.,\ $c \leq (n-1) \Gamma_{_{\hspace{-1pt}\textrm{Tx}}}$, Algorithm \ref{alg:Chunking} 
	would add more transactions to the current chunk (line \ref{alg:Chunking:add_decission}),
	as 	the computation consumption for transactions is upper-bounded by $\Gamma_{_{\hspace{-1pt}\textrm{Tx}}}$.
	\item 
	$c \leq n \Gamma_{_{\hspace{-1pt}\textrm{Tx}}}$, which is guaranteed by Algorithm \ref{alg:Chunking}, line \ref{alg:Chunking:add_decission}.
\end{itemize}
Hence, all chunks, except for the last one, will have a computation consumption $c$ 
\begin{align}\label{eq:ExecutionFlow:Execution:Chunking:ComputationConsumption}
	(1 - \tfrac{1}{n}) \Gamma_{_{\hspace{-1pt}\textrm{chunk}}} < c \leq \Gamma_{_{\hspace{-1pt}\textrm{chunk}}}.
\end{align}
Choosing $n$ large enough guarantees that all but the last chunk have similar computation consumption. 

The last chunk could contain as little as a single transaction. 
Hence, its computation consumption $c_{_\textnormal{LastChunk}}$ could take any value 
$0 < c_{_\textnormal{LastChunk}} \le \Gamma_{_{\hspace{-1pt}\textrm{chunk}}}$.
As opposed to $(1 - \nicefrac{1}{n}) \Gamma_{_{\hspace{-1pt}\textrm{chunk}}} < c $ for any other chunk in the block.
The last chunk being significantly smaller than $(1 - \nicefrac{1}{n}) \Gamma_{_{\hspace{-1pt}\textrm{chunk}}}$ 
does not pose a problem for the following reason. 
For example, consider a node participating in a network that can process blocks with up to $\Xi = 1000$ chunks.
For  $\eta=4.2\%$, a node would be required to have the capacity to process up to $42$ chunks per block (as outlined above). 
Hence, for each block with $\Xi' \le 42$, an honest Verifier would simply check the entire block. 
For blocks with more chunks, the workload is failry uniform across all Verifiers, 
even though each Verifier samples a subset of chunks for verification. 
This is because the large majority of all chunks have fairly comparable computation consumption 
as derived in equation \eqref{eq:ExecutionFlow:Execution:Chunking:ComputationConsumption}.

\subsubsection{The Execution Receipt\label{sec:ExecutionFlow:Execution:ExecutionReceipt}}

As Message \ref{msg:ExecutionReceipt} shows, the \techterm{Execution Receipt}  is a signed \texttt{ExecutionResult}, 
which provides authentication of the sender and guarantees integrity of the message.  
The \texttt{ExecutionResult} encapsulates the Execution Node's commitments to both its interim and final results.  
Specifically, it contains a reference, \texttt{ExecutionResult.blockHash},  
to the block whose result it contains and \texttt{ExecutionResult.finalState} as a commitment%
\footnote{%
	We don't specify details of the \texttt{StateCommitment} here. It could either be the full state (which is likely much too large for a mature system). 
	Alternatively,  \texttt{StateCommitment} could be the output of a hashing algorithm, such as \techterm{Merkle Patricia Trie} used by Etherum  \cite{Ethereum:PatriciaTree}.
}
to the resulting state.

\begin{wrapfigure}{R}{0.45\textwidth}
	\begin{minipage}{0.45\textwidth}
		\vspace{-20pt}
		\begin{algorithm}[H]
			\renewcommand{\thealgorithm}{}
			\floatname{algorithm}{}
			\begin{algorithmic}[1]
				\State \textbf{message} Chunk $\{$
				\State \quad \textbf{StateCommitment}  startState;
				\State \quad \textbf{float}  $\tau_0$;
				\State \quad \textbf{uint32}  startingTransactionIndex;
				\State \quad \textbf{float}  computationConsumption;
				\State $\}$\vspace{5pt}
			\end{algorithmic}
			\hrule
			\begin{algorithmic}[1]
				\State \textbf{message} ExecutionResult $\{$
				% \State \quad \textbf{uint64} blockHeight;
				\State \quad \textbf{bytes} blockHash;
				\State \quad \textbf{bytes} previousExecutionResultHash;
				\State \quad \textbf{repeated Chunk} chunks; 
				\State \quad \textbf{StateCommitment}  finalState;
				\State $\}$\vspace{5pt}
			\end{algorithmic}		
			\hrule
			\begin{algorithmic}[1]
				\State \textbf{message} ExecutionReceipt $\{$
				\State \quad \textbf{ExecutionResult} executionResult;
				\State \quad \textbf{repeated SPoCK} \textit{Zs};
				\State \quad \textbf{bytes} executorSig;
				\State $\}$
			\end{algorithmic}
		\end{algorithm}
	\end{minipage}
	\captionof{messagedef}{
		Execution Nodes broadcast an \texttt{ExecutionReceipt} to the entire network after computing a block.
	}
	\label{msg:ExecutionReceipt}
	\vspace{-29pt}
\end{wrapfigure}

Consider the situation where an Execution Node truthfully computed a block but used the computational state from a faulty Execution Receipt as input. 
While the node's output will likely propagate the error,
it is important to attribute the  error to the malicious node that originally introduced it and slash the malicious node. 
To ensure attributability of errors,  \texttt{ExecutionResult}s specify the computation result they built on top of as \texttt{ExecutionResult.previousExecutionResultHash}. 
\\
As we will show later in section \ref{sec:ExecutionFlow:CorrectnessProof}, a faulty computation result will be rejected with overwhelming probability. 

Lastly, the \texttt{ExecutionResult.chunks} contains the Execution Node's result 
of running the \texttt{Chunking} algorithm 
(\texttt{Chunk.startingTransactionIndex} and \\
\texttt{Chunk.computationConsumption}).
Also, a \texttt{Chunk} states the starting state (\texttt{Chunk.startState}) for the computation and the 
computation consumption for the first transaction in the chunk (\texttt{Chunk.}$\tau_0$)

\subsubsection*{Solving the Verifier's Dilemma}

The field \texttt{ExecutionReceipt.}\textit{Zs} is a list of \hyperref[sec:Z]{\techterm{SPoCK}s}  which is generated based on interim states from computing the  chunks. 
Formally, for the $i^\textnormal{\,th}$ chunk 
\texttt{VerificationProof.}\textit{Zs}$[i]$ holds the \techterm{SPoCK} demonstrating knowledge of the secret derived from executing the chunk. 
As explained in section \ref{sec:ExecutionFlow:Verification:VerificationByRe-Computation}, 
the \techterm{SPoCK} is required to resolve the  \techterm{Verifier's Dilemma} in \Architecture.
Furthermore, it prevents Execution Nodes from copying \texttt{ExecutionResult} from each other,  pretending their computation is faster than it is.

Given that there are several Execution Nodes, it is likely that multiple Execution Receipts are issued for the same block. We define \techterm{consistency} of Execution Receipts as follows. 

\begin{definition}\label{def:ExecutionReceiptConsistency}\textit{Consistency Property of Execution Receipts}\\
	Execution Receipts are \textbf{consistent} if and only if  
	\begin{enumerate}
		\item their \texttt{ExecutionResult}s are identical and
		\item their \hyperref[sec:Z]{\techterm{SPoCK}s} attest to the same confidential knowledge.
	\end{enumerate}
\end{definition}

\subsection{Verification Role: Checking Execution Result\label{sec:ExecutionFlow:Verification}}

\noindent
The verification process is designed to probabilistically guarantee safety of the computation result 
(see page \pageref{theorem:ExecutionFlow:CorrectnessProof}, Theorem \ref{theorem:ExecutionFlow:CorrectnessProof}). 
A crucial aspect for the computational safety is that Verification Nodes verifiably \emph{self-select} 
the chunks they check \emph{independently} from each other. This process is described in the section below. 
In section \ref{sec:ExecutionFlow:Verification:VerificationByRe-Computation}, we describe how the selected chunks are verified. 

\subsubsection{Self-selecting chunks for verification\label{sec:ExecutionFlow:Verification:SelfSelection}}

The protocol by which Verification Nodes \emph{self-select} their chunks is given in Algorithm \ref{alg:ChunkSelection}. 
While inspired by Algorand's \techterm{random sortition} \cite{Algorand:2017:ScalingByzantineAgreementsForCryptocurrencies}, 
it has significantly reduced complexity. 
\medskip

\newpage
\noindent
\texttt{ChunkSelfSelection} has the following crucial properties.
\begin{enumerate}[label=(\arabic*)]
	\item \textit{Verifiability}.\label{item:ExecutionFlow:Verification:SelfSelection:Verifiability}
	The seed for the random selection of the chunks is generated in  lines \ref{alg:ChunkSelection:proof} -- \ref{alg:ChunkSelection:seed}.
	Given the \texttt{ExecutionReceipt}, the Verifier's public key, and the proof $p$, any other party can confirm that $p$ was generated according to protocol. 
	Moreover, given $p$, anyone can re-compute the Verifier's expected chunk assignment. 
	Thereby, the chunk-selection protocol is verifiable and deviating from the protocol is detectable, attributable, and punishable. 
	\item \textit{Independence}.\label{item:ExecutionFlow:Verification:SelfSelection:Independence}
	Each Verifier samples its chunks locally and independently from all the other Verifiers.  
	\item \textit{Unpredictability}. 
	Without knowledge of the Verifier's secret key \texttt{sk}, it is computationally infeasible to predict the sample. 
	Formally, the computation of \texttt{seed} can be considered a \techterm{verifiable random function} \cite{Micali:1999:VRFs}. 
	\item \textit{Uniformity}. 	
	A Verifier uses \techterm{Fisher-Yates} shuffling \cite{FisherYates:1974:Shuffling, knuth:1997:SeminumericalAlgorithms} to self-select the chunks it checks.
	The \techterm{Fisher-Yates} algorithm's pseudo-random number generator is seeded with the
	output of a cryptographic hash function.
	The uniformity of the seed and the uniformity of the shuffling algorithm together guarantee the uniformity of the generated chunk selection. 
\end{enumerate}
\bigskip

\begin{figure}[t!]
\raggedleft
\vspace{-10pt}
\begin{algorithm}[H]
	\caption{ChunkSelfSelection}
	\textbf{Input:} 
	\begin{itemize}
		\item[$\eta$:] fraction of chunks to be checked by a Verification Node
		\item[\texttt{er}:] Execution Receipt
		\item[\texttt{sk}:] Verifier Node's secret key 
	\end{itemize}
	\textbf{Output:}
	\begin{itemize}
		\item[$L$:] List of chunk indices that are assigned to the Verifier 
		\item[$p$:] proof of protocol-compliant selection
	\end{itemize}			
	\begin{algorithmic}[1]
		\State \textbf{ChunkSelfSelection}($\eta$, \texttt{er}, \texttt{sk}) 
		\State \quad $\texttt{chunks} \leftarrow \texttt{er.executionResult.chunks}$ \Comment{list of chunks from Execution Receipt}
		\State \quad $\Xi \leftarrow \texttt{length}(\texttt{chunks})$ \Comment{number of chunks in the block}		 
		\State \quad $p \leftarrow \texttt{sign}_\texttt{sk}(\texttt{er.executionResult})$ \label{alg:ChunkSelection:proof}  \Comment{sign Execution Receipt's \techterm{ExecutionResult}}	
		\State \quad $\texttt{seed} \leftarrow \texttt{hash}(p)$  \label{alg:ChunkSelection:seed} \Comment{use signature's hash as random number generator seed}
		\State \quad $n \leftarrow \ceil*{\eta \cdot \Xi}$ \label{alg:ChunkSelection:numberElements} \Comment{number of chunks for verification}
		\State \quad $L \leftarrow$ \texttt{FisherYatesShuffle}(\texttt{chunks}, \texttt{seed}, $n$) \label{alg:ChunkSelection:FY} \Comment{generate random permutation of chunks}
		\State \quad \textbf{return} $L, p$  
	\end{algorithmic}
\end{algorithm}
\vspace{-15pt}
\captionof{algorithmdef}{
	randomly selects chunks for subsequent verification. 
	In line \ref{alg:ChunkSelection:numberElements}, $\ceil*{\cdot}$ is the ceiling operator. 
	The function  \texttt{FisherYatesShuffle}(\texttt{list}, \texttt{seed}, $n$) draws a simple random sample without replacement
    from the input \texttt{list}. 
    The \texttt{seed} is used for initializing the pseudo-random-number generator. 
}
\label{alg:ChunkSelection}
\end{figure}

\begin{corollary}\label{cor:Chunk-Self-selection-Properties}$~$\\
	Algorithm \ref{alg:ChunkSelection} samples a subset of  $\ceil*{\eta \cdot \Xi}$ chunks, 
	for $\Xi$ the number of chunks in the block and $\eta$ the fraction of chunks to be checked by each Verification Node. 
	The selection probability \emph{uniform} and 
	\emph{independently and identically distributed} (i.i.d.) for all Verifiers.  
	The sample is \emph{unpredictable} without the Verification Node's secret key. 
\end{corollary}
Independence and unpredictability are crucial for the system's security. 
A selection protocol without these properties might be abused by a malicious Execution Node (see section \ref{sec:attack-vectors:ColludingENandVNs} for a detailed discussion). 

Our approach to independent verification of chunks has similarities with traditional \techterm{acceptance sampling} theory 
\cite{Wetherill1977:AcceptanceSamplingBasicIdea, SchillingNeubauer:2017AcceptanceSamplingInQualityControl, Haakan:2002:ProbabilisticVerification}, 
yet differs as our model assumptions are different.
In contrast to traditional acceptance sampling, where physical items are tested, identical copies of our digital chunks can be checked in parallel by multiple Verifiers.
Part of the novelty of our approach is that we elevate an acceptance sampling model with parallel sampling. 

\subsubsection{Verifying chunks\label{sec:ExecutionFlow:Verification:VerificationByRe-Computation}}

The verification protocol is designed to be self-contained, 
meaning any \techterm{Execution Receipt} can be verified in isolation. All required data is specified through the hashes in the execution receipt.
Checking the correctness of a chunk requires re-computing the transactions in the chunk. 
The details of the verification protocol are given in Algorithm \ref{alg:ChunkVerification}. 
The inputs $\tau_0$ and $\tau_0'$ for the \texttt{ChunkVerification} algorithm are taken directly from the \techterm{Execution Receipt}. 
$\Lambda$, \texttt{Txs},  and $\Lambda'$,  have to be fetched from the Execution Nodes 
and checked to match the  \techterm{Execution Receipt} (specifically: $\Lambda$, $\Lambda'$) or the original \techterm{block} (specifically: \texttt{Txs}) via hashing%
\footnote{%
	We use the term `hashing' here to refer to the one-time application of a conventional hash function as well as iterative hashing schemes such as Merkle trees or Merkle Patricia trie.
}.
Therefore, errors uncovered by the verification process can be attributed to the data provider and slashed.

\begin{figure}[t!]
	\raggedleft
	\vspace{-10pt}
	\begin{algorithm}[H]
		\caption{ChunkVerification}
		\textbf{Input:}  \hspace{35.5pt}$\Lambda$: starting state for computing the chunk
		\begin{itemize}[leftmargin=90.5pt]
			\item[\texttt{Txs}:] List of transactions in chunk
			\item[$\tau_0$:] computation consumption of leading transaction in \emph{this} chunk
			\item[$\Lambda'$:] resulting state after computing the chunk as stated in the \techterm{Execution Receipt}
			\item[$\tau_0'$:] computation consumption of leading transaction in \emph{next} chunk; \\ or $\infty$ if this is the last chunk in the block
			\item[$c$:] chunk's computation consumption as stated in the \techterm{Execution Receipt}
		\end{itemize}
		\textbf{Output:} \quad \texttt{true}: if and only if chunk passes verification
		\begin{algorithmic}[1]
			\State \textbf{ChunkVerification}($\Lambda$, \texttt{Txs}, $\Lambda'$, \texttt{ntx}, \texttt{c}) 
			\State \quad $\gamma \leftarrow 0$ \Comment{{\footnotesize accumulated computation consumption}}
			\State \quad \textbf{for} $t$ \textbf{in} \texttt{Txs}: \Comment{{\footnotesize for each transaction in chunk}}
			\State \quad \quad $\Lambda, \tau \leftarrow $ \texttt{execute}($\Lambda$, $t$)  \Comment{{\footnotesize execute transaction}}
			\State \quad \quad \texttt{if} $t$ is first transaction in chunk:
			\State \quad \quad \quad \textbf{assert that}  $\tau_0 ==  \tau$ \Comment{{\footnotesize computation consumption for first transaction in chunk is correct}}
			\State \quad \quad $\gamma \leftarrow \gamma + \tau$ \Comment{{\footnotesize add transaction's computation consumption to $\gamma$}}
			\State \quad \textbf{assert that} $\gamma == c$ \Comment{{\footnotesize computation consumption for entire chunk is correct}}
			\State \quad \textbf{assert that} $c \leq \Gamma_{_{\hspace{-1pt}\textrm{chunk}}}$ \Comment{{\footnotesize computation consumption does not exceed limit}}
			\State \quad \textbf{assert that} $c + \tau_0' > \Gamma_{_{\hspace{-1pt}\textrm{chunk}}}$ \Comment{{\footnotesize chunk is full: no more translations can be appended to chunk}}
			\State \quad \textbf{assert that} $\Lambda == \Lambda'$ \Comment{{\footnotesize verify Execution Node's resulting state}}
			\State \quad \textbf{return} \texttt{true} 
		\end{algorithmic}
	\end{algorithm}
	\vspace{-13pt}
	\captionof{algorithmdef}{
		verifies a chunk. The function \texttt{execute}($\Lambda$, $t$) applies the transaction $t$ to the computational state $\Lambda$ and returns a pair of values:
		the resulting state (first return value) and transaction's computation consumption (second return value). The 
		\texttt{assert} statement raises an exception if the condition is false.
	}
	\label{alg:ChunkVerification}
	\vspace{5pt}	
\end{figure}

\begin{figure}[t!]
	\centering
	\vspace{-10pt}
	\begin{minipage}{1.0\textwidth}
		\begin{algorithm}[H]
			\renewcommand{\thealgorithm}{}
			\floatname{algorithm}{}
			\begin{algorithmic}[1]
				\State \textbf{message} CorrectnessAttestation $\{$
				\State \quad \textbf{bytes} executionResultHash; \Comment{{\footnotesize Hash of  approved \texttt{ExecutionResult}}}
				\State \quad \textbf{bytes} attestationSig; \Comment{{\footnotesize Signature over \texttt{executionResultHash}}}
				\State $\}$
			\end{algorithmic}
			\hrule
			\begin{algorithmic}[1]
				\State \textbf{message} VerificationProof $\{$
				\State \quad \textbf{repeated uint32} $L$; \Comment{{\footnotesize list of chunk indices assigned to verifer}}
				\State \quad \textbf{bytes} $p$; \Comment{{\footnotesize proof to verify correctness of chunk assignment}}
				\State \quad \textbf{repeated SPoCK} \textit{Zs}; \Comment{{\footnotesize for each assigned chunk: proof of re-computation}}
				\State $\}$
			\end{algorithmic}
			\hrule			
			\begin{algorithmic}[1]
				\State \textbf{message} ResultApproval $\{$
				% \State \quad \textbf{uint64} blockHeight;
				\State \quad \textbf{CorrectnessAttestation} attestation;
				\State \quad \textbf{VerificationProof} verificationProof; 
				\State \quad \textbf{bytes} verifierSig; \Comment{{\footnotesize signature over all the above fields}}
				\State $\}$
			\end{algorithmic}
		\end{algorithm}
		\vspace{-13pt}
	\end{minipage}
	\captionof{messagedef}{
		Verification Nodes broadcast a \texttt{ResultApproval} to all Consensus Nodes if all their assigned chunks pass verification.
	}
	\label{msg:ResultApproval}
\end{figure}

The verification process is also given enforcement power, as we enable it to request slashing against an Execution Node.
A successful verification process results in a \texttt{ResultApproval} (Message \ref{msg:ResultApproval})  being broadcast by the Verifier to all Consensus Nodes. 
It is important to note that a \texttt{ResultApproval} (specifically \texttt{ResultApproval.attestation}) attests to the correctness of an \texttt{ExecutionResult}. 
Specifically, in \texttt{CorrectnessAttestation}, the Verifier signs the \texttt{ExecutionResult}, \emph{not} the \techterm{Execution Receipt}.
Per definition \ref{def:ExecutionReceiptConsistency}, multiple \emph{consistent} \techterm{Execution Receipts} have identical \texttt{ExecutionResult}s. 
Hence, their correctness is simultaneously attested to by  a single \texttt{ResultApproval} message. 
This saves communication bandwidth, as each Verifier Node issues only one 
\texttt{ResultApproval} for the common case that several Execution Nodes publish the same results by issuing consistent \techterm{Execution Receipts}.

The second important component is the \texttt{ResultApproval.verificationProof}, which proves that the Verification Node completed the assigned verification tasks. 
We designed this protocol component to address the \techterm{Verifier's Dilemma} \cite{Luu:2015:VerifierDilemma}. 
It prevents the Verifier from just approving \techterm{ExecutionReceipts}, betting on honesty of the Execution Nodes, 
and thereby being compensated for work the Vertifier didn't do. 
The field \texttt{VerificationProof.}$L$ specifies the chunk indices the Verifier has selected by running \texttt{ChunkSelfSelection} (Algorithm \ref{alg:ChunkSelection}). 
As detailed in section \ref{sec:ExecutionFlow:Verification:SelfSelection} 
(property \hyperref[item:ExecutionFlow:Verification:SelfSelection:Verifiability]{\ref{item:ExecutionFlow:Verification:SelfSelection:Verifiability} \textit{Verifiability}}),
protocol-compliant chunk selection is proven by \texttt{VerificationProof.}$p$, which holds the value for $p$ returned by \texttt{ChunkSelfSelection}. 
The list \texttt{VerificationProof.}\textit{Zs} contains for each assigned chunk a proof of verification. 
Formally, for each $i \in \texttt{VerificationProof}.L$, \texttt{VerificationProof.}\textit{Zs}$[i]$ holds a \hyperref[sec:Z]{\techterm{SPoCK}}. 

Each Verifier samples the chunks it checks independently 
(property \ref{item:ExecutionFlow:Verification:SelfSelection:Independence}).
Furthermore, statistically each chunk is checked by a larger number of nodes (e.g.,\ on the order of 40 as suggested in section \ref{sec:ExecutionFlow:CompLoadOnVerifier})
Therefore, with overwhelming probability,  all chunks are checked.  (We formalize this argument in Theorem \ref{theorem:ExecutionFlow:CorrectnessProof}).

\subsection{Consensus Role: Sealing Computation Result\label{sec:ExecutionFlow:BlockSealing}}

\begin{figure}[b!]
	% figure source: https://docs.google.com/drawings/d/16IvANReeELQ2GBqD6R2cM7kFZ8Do_8EW57axjBtlDcc/edit?usp=sharing_eip&ts=5d2448a6
	\centering
	\includegraphics[trim=0 0 0 30, clip, width=\textwidth]{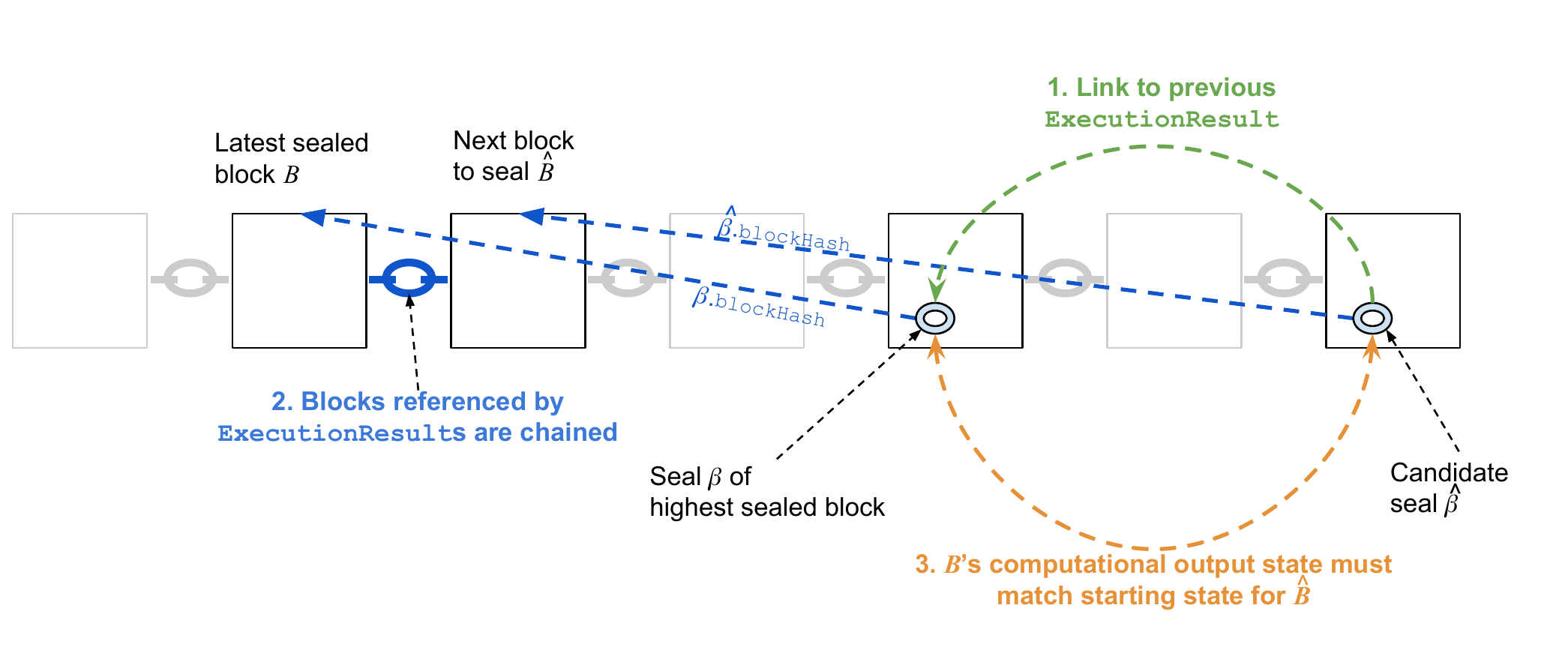}
	\caption{
		Validity conditions for a new BlockSeal according to Algorithm \ref{alg:BlockSealValidity}.
	}
	\label{fig:BlockSealing}
\end{figure}

\noindent
\techterm{Sealing} a block's computation result happens after the block itself is already finalized.
Once the computation results have been broadcast as \techterm{Execution Receipts}, 
Consensus Nodes wait for the \texttt{ExecutionResult}s to accumulate matching \techterm{Results Approvals}.
Only after a super-majority of Verifier Nodes approved the result 
and \emph{no} \texttt{FaultyComputationChallenge} has been submitted (see section \ref{sec:AttackMitigation:FaultyComputationResult} for details), 
the  \texttt{ExecutionResult} is considered for sealing by the Consensus Nodes. 
The content of a \texttt{BlockSeal} is given in Message \ref{msg:BlockSeal}.
Algorithm \ref{alg:BlockSealValidity} specifies the full set of conditions a \texttt{BlockSeal} must satisfy.
Algorithm \ref{alg:BlockSealValidity} enforces a specific structure in our blockchain which is illustrated in Figure \ref{fig:BlockSealing}.
\medskip

\noindent
Once a Consensus Node finds that all conditions are valid, 
it incorporates the   \texttt{BlockSeal} as an element of \texttt{Block.blockSeals} into the next Block it proposes. 
All honest Consensus Nodes will check the validity of the \texttt{BlockSeal} as part of verifying the Block, \emph{before} voting for it.
Thereby the validity of a \texttt{BlockSeal} is guaranteed by the BFT consensus algorithm. 
Furthermore, condition \ref{alg:blockSealValidity_cond:waiting} guarantees that 
any  \texttt{FaultyComputationChallenge}  has been received \emph{before} the block is sealed. 
This gives rise to the following corollary:
\medskip

\begin{corollary}\label{corollary:blocksealing}$~$\\
	Given a system with partially synchronous network conditions with message traversal time bounded by $\Delta_t$.
	A Block is only sealed if more than $\frac{2}{3}$ of Verification Nodes approved the \texttt{ExecutionResult} and no  \texttt{FaultyComputationChallenge} was submitted and  adjudicated with the result that 
	the  \texttt{ExecutionResult} is faulty.
\end{corollary}

\newpage
\begin{figure}[h!]
	\centering
	\vspace{-10pt}
	\begin{minipage}{1.0\textwidth}
		\begin{algorithm}[H]
			\renewcommand{\thealgorithm}{}
			\floatname{algorithm}{}
			\begin{algorithmic}[1]
				\State \textbf{message} BlockSeal $\{$
				\State \quad  \textbf{bytes} blockHash;
				\State \quad  \textbf{bytes} executionResultHash;
				\State \quad  \textbf{bytes} executorSigs; \Comment{{\footnotesize Signatures from Execution Nodes over \texttt{ExecutionResult}}}
				\State \quad  \textbf{bytes} attestationSigs; \Comment{{\footnotesize Signatures from Verification Nodes approving the \texttt{ExecutionResult}}}
				\State \quad  \textbf{bytes} proofOfWaiting; \Comment{{\footnotesize output of VDF to prove waiting for a  \texttt{FaultyComputationChallenge} }}
				\State $\}$
			\end{algorithmic}
		\end{algorithm}
		\vspace{-5pt}
		\captionof{messagedef}{
			In order to seal a block (with hash \texttt{blockHash}), Consensus Nodes add a \texttt{BlockSeal} to the next block  (field \hyperref[msg:Block]{\texttt{Block}}\texttt{.blockSeals}) they finalize.
			The field \texttt{executorSigs} is the aggregated signature of at least one Execution Node that published an Execution Receipt with a compatible 
			\texttt{ExecutionResult}.
			Their \texttt{BlockSeal} is valid only if more than $\frac{2}{3}$ of Verifier  Nodes have signed. 
			Instead of storing the signatures individually, we use store an aggregated signature in \texttt{attestationSigs}.  	
		}
		\label{msg:BlockSeal}
		\vspace{15pt}		
	\end{minipage}
\end{figure}

\begin{figure}[h!]
	\raggedleft
	\vspace{-10pt}
	\begin{algorithm}[H]
		\caption{Validity of Block Seal}
		Let $\beta$ be the (accepted) \texttt{BlockSeal} for the \emph{highest} sealed block, i.e.,\ $\beta$  is contained in a finalized block. 
		A candidate \techterm{BlockSeal} $\hat{\beta}$ must satisfy the following conditions to be valid.
		\begin{enumerate}[label=(\arabic*), leftmargin=*]
			\item \label{alg:blockSealValidity_cond_1}
			$\hat{\beta}$\texttt{.executionResult.previousExecutionResultHash} must be equal to\\ $\beta$\texttt{.executionResultHash}
			\item 
			Let 
			\begin{itemize}
				\item $B$ be the block whose result is sealed by $\beta$, i.e.,\ $B$ is referenced by $\beta$\texttt{.blockHash};
				\item $\hat{B}$ be the block whose result is sealed by $\hat{\beta}$, i.e.,\ $\hat{B}$ is referenced by $\hat{\beta}$\texttt{.blockHash}.
			\end{itemize}
			$\hat{B}$\texttt{.previousBlockHash} must reference $B$.
			\item \label{alg:blockSealValidity_cond_2}
			Let 
			\begin{itemize}
				\item $E$ be the \texttt{ExecutionResult} that referenced (sealed) by $\beta$\texttt{.executionResultHash};
				\item $\hat{E}$ be the \texttt{ExecutionResult} that is referenced by $\hat{\beta}$\texttt{.executionResultHash}.
			\end{itemize}
			The starting state for computing $\hat{B}$ must match the $B$ computational output state, i.e.,\
			\begin{align*}
			\hat{E}\texttt{.chunks}[0]\texttt{.startState} == E\texttt{.finalState}
			\end{align*}
			\item \label{alg:blockSealValidity_cond_3}
			$\hat{\beta}$\texttt{.attestationSigs} must contain more than $\frac{2}{3}$ of Verifier  Nodes' signatures 
			\item
			For each Verifier who contributed to $\hat{\beta}$\texttt{.attestationSigs}, the \texttt{VerificationProof} has been validated. 
			\item \label{alg:blockSealValidity_cond:pendingFCC}
			No \texttt{FaultyComputationChallenge} against the \texttt{ExecutionResult}  is  \emph{pending} (i.e.,\ not yet adjudicated).
			\item \label{alg:blockSealValidity_cond:noApprovedFCC}
			No \texttt{FaultyComputationChallenge} against the \texttt{ExecutionResult} was adjudicated with the result that 
			the  \texttt{ExecutionResult} is faulty.
			\item  \label{alg:blockSealValidity_cond:waiting}
			$\beta$\texttt{.proofOfWaiting} proves a sufficiently long wait time $\Delta_t$
		\end{enumerate}
		Per axiom, we consider the genesis block as sealed. 
	\end{algorithm}
	\vspace{-15pt}
	\captionof{algorithmdef}{
		protocol for determining validity of a \texttt{BlockSeal}. 
		Figure \ref{fig:BlockSealing} illustrates conditions \ref{alg:blockSealValidity_cond_1} -- \ref{alg:blockSealValidity_cond_3}.
	}
	\label{alg:BlockSealValidity}
	\vspace{10pt}	
\end{figure}

\newpage
\subsection{Proof of Computational Correctness\label{sec:ExecutionFlow:CorrectnessProof}}

Below, we prove in Theorem \ref{theorem:ExecutionFlow:LivenessProof} that 
\Architecture's computation infrastructure has guaranteed liveness,
even in the presence of a moderate number of Byzantine actors. 
Furthermore, Theorem \ref{theorem:ExecutionFlow:CorrectnessProof} proves that block computation is \emph{safe},
i.e.,\ the resulting states in sealed blocks are correct with overwhelming probability. 
While safety is unconditional on the network conditions, liveness requires a partially synchronous network.

\begin{theorem}\label{theorem:ExecutionFlow:LivenessProof}
	\textit{\textbf{Computational Liveness}}\\
	Given a system with
	\begin{itemize}
		\item more than $\frac{2}{3}$ of the Consensus Nodes' stake is controlled by honest actors;
		\item and at least one honest Execution Node; 
		\item and more than $\frac{2}{3}$ of the Verification Nodes' stake is controlled by honest actors; 
		\item and partially synchronous network conditions with message traversal time bounded by $\Delta_t$.
	\end{itemize}
	The computation and sealing of finalized blocks always progresses.
\end{theorem}

\noindent%
\textbf{Proof of Theorem  \ref{theorem:ExecutionFlow:LivenessProof}} 
\begin{itemize}
	\item 
	Assuming liveness of the consensus algorithm, finalization of new blocks always progresses.
	\item 
	For a system with one honest Execution Node, there is at least one \techterm{Execution Receipt} with a correct \texttt{ExecutionResult}.
	\item 
	Every honest Validator will approve a correct \texttt{ExecutionResult}. Hence, there will be \techterm{Result Approvals} by at least $\frac{2}{3}$ of the Verification Nodes. 
	\item 
	Malicious Verifiers might temporarily delay block sealing by raising a \texttt{FaultyComputation}-\texttt{Challenge},  which triggers condition \ref{alg:blockSealValidity_cond:pendingFCC} in Algorithm  \ref{alg:BlockSealValidity}. However, the resolution process (see section \ref{sec:AttackMitigation:FaultyComputationResult} for details)
	guarantees that the \texttt{FaultyComputationChallenge} is eventually adjudicated and malicious Verifiers are slashed (Corollary  \ref{corollary:AttackMitigation:FaultyComputationResult}). 
	Therefore, malicious Verifiers cannot indefinitely suppress 
	block sealing via condition \ref{alg:blockSealValidity_cond:pendingFCC}
	or even reach condition \ref{alg:blockSealValidity_cond:noApprovedFCC}.
	\item 
	Consequently, all honest Consensus nodes will eventually agree on the validity of the \techterm{Block Seal}.
	\item 
	Assuming a consensus algorithm with guaranteed \techterm{chain quality}\footnote{
		Formally, \techterm{chain quality} of a blockchain is the ratio 
		$\frac{\textnormal{\techterm{number of blocks contributed by honest nodes}}}{\textnormal{\techterm{number of blocks contributed by malicious nodes}}}$
		\cite{Garay:2015:ChainQuality}.
	},
	an honest Consensus Node will eventually propose a block and include the \techterm{Block Seal} as prescribed by the protocol. 
	\item 
	Given that there are more than $\frac{2}{3}$ of honest Consensus Nodes, the block containing the seal will eventually be finalized (by liveness of consensus). 
\end{itemize}
\vspace{-15pt}
\begin{flushright}
	$\square$	
\end{flushright}	

\begin{theorem}\label{theorem:ExecutionFlow:CorrectnessProof}
	\textit{\textbf{Computational Safety}}\\
	Given a system with
	\begin{itemize}
		\item partially synchronous network conditions with message traversal time bounded by $\Delta_t$;
		\item more than  $\frac{2}{3}$ of the Consensus Nodes' stake is controlled by honest actors;
		\item all Verification Nodes are equally staked and more than $\frac{2}{3}$ of them are honest.
	\end{itemize}
    Let $\widetilde{N}$ denote the number of honest Verification Nodes 
    and $\eta$ the  \emph{fraction} of chunks each Verifier checks. 
	The probability  $P_\textrm{error}$ for a computational error in a sealed block is bounded by 
	\begin{align}\label{theorem:ExecutionFlow:CorrectnessProof:block_error_probability}
		P_\textrm{error} \lesssim \Xi \cdot  \big(1 - \eta\big)^{\widetilde{N}}
	\end{align}
	for large $\widetilde{N}$. Here, $\Xi$ denotes the number of chunks in the \techterm{Execution Receipt}. 
\end{theorem}

\noindent
Theorem  \ref{theorem:ExecutionFlow:CorrectnessProof} states that the probability of a computational error decreases \emph{exponentially} with the number of Verifiers. 
\medskip 

\noindent%
\textbf{Proof of Theorem  \ref{theorem:ExecutionFlow:CorrectnessProof}} \\
Consider an  \techterm{Execution Receipt}.  
\begin{itemize}
	\item 
	For brevity, we denote the \techterm{Execution Receipt}'s chunks as $L$ , i.e.,\ 
	\[ L =  \texttt{ExecutionReceipt.executionResult.chunks}. \]
	Without loss of generality, we treat $L$  as a \emph{set} (instead of an ordered list), as this proof does not depend on the order of $L$'s elements. 
	\item Let $\Xi$ be the total number of chunks in the \techterm{Execution Receipt}, i.e\ $\Xi = |L|$, where $|\cdot|$ denotes the cardinality operator. 
	\item Let $\xi \in L$ denote a chunk.
\end{itemize}

\noindent
As formalized in Corollary \ref{cor:Chunk-Self-selection-Properties}, each honest Verifier randomly selects a subset $S \subset L$ with $|S| = \ceil*{\eta \cdot \Xi}$ 
by execution \texttt{ChunkSelfSelection} (Algorithm \ref{alg:ChunkSelection}). 
As each Verifier selects the chunks by Fisher-Yates shuffle, it follows that chunk $\xi$ not being selected as the first element is $\frac{\Xi -1}{\Xi}$ 
and that $\xi$ is not selected as the second element is $\frac{\Xi -2}{\Xi-1}$, etc. 
Hence, the probability that chunk $\xi$  is \emph{not checked} by \emph{one specific verifer} is
\begin{align}
	\bar{p}_\xi = \frac{\Xi -1}{\Xi} \cdot \frac{\Xi -2}{\Xi-1} \cdot \ldots \cdot \frac{\Xi - n}{\Xi- (n-1)} = \frac{\Xi - n }{\Xi} = 1 - \frac{n}{\Xi}, \qquad \textnormal{for } n := |S| = \ceil*{\eta \cdot \Xi}.
\end{align}
Let $\bar{P}_\xi$ be the probability that $\xi$ is checked by \emph{none} of the honest Verifiers.
\begin{align}\label{theorem:ExecutionFlow:CorrectnessProof:allchunkschecked}
	\bar{P}_\xi  = \bar{p}_\xi^{\widetilde{N}} =  &\big(1 - \tfrac{n}{\Xi}\big)^{\widetilde{N}} \\
	                  \leq &\big(1 - \eta\big)^{\widetilde{N}} 
\label{theorem:ExecutionFlow:CorrectnessProof:numberChunksToBeChecked}	
\end{align}
as $n = \ceil*{\eta \cdot \Xi} \geq \eta \cdot \Xi$. 
Consequently, the probability of \emph{the specific} chunk $\xi$ not being checked decreases \emph{exponentially} with the number of Verifiers $\widetilde{N}$. 
\begin{figure}[t!]
	\centering
	\includegraphics[trim=0 0 0 30, clip, width=0.6\textwidth]{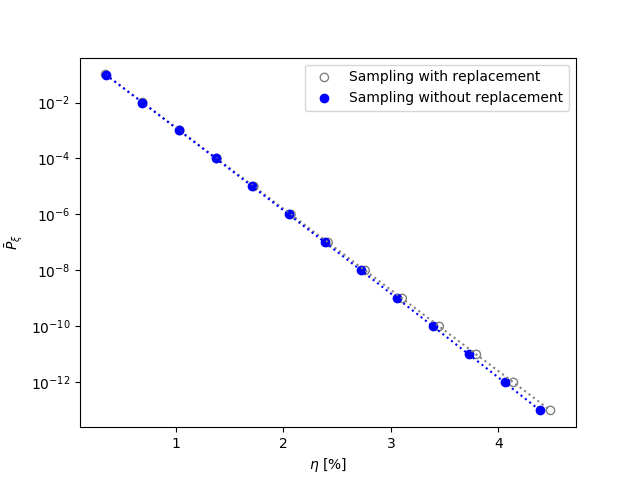}
	\caption{
		Probability $\bar{P}_\xi$ that a specific chunk is checked by \emph{none} of the honest Verifiers. 
		$\eta$ is the \emph{fraction} of chunks each Verifier selects for checking. The graph illustrates 
		probabilities for $\widetilde{N}=667$ honest Verification Nodes verifying an \techterm{Execution Receipt} 
		with $\Xi = 1000$ chunks. The blue curve shows the dependency when Verifiers sample
		their chunks based on Fisher-Yates as specified in Algorithm  \ref{alg:ChunkSelection}, i.e.,\
		sample chunks from the \techterm{Execution Receipt}  \emph{without replacement}.
		For comparison, we show sampling with replacement.
	}
	\label{fig:Prob_Sealing_Wrong_Result}
\end{figure}
Figure \ref{fig:Prob_Sealing_Wrong_Result} visualizes the \emph{exact} probability $\bar{P}_\xi$ as a function of $\eta$  
as given in equation \eqref{theorem:ExecutionFlow:CorrectnessProof:allchunkschecked}. 

The probability that \emph{all} chunks are checked by \emph{at least} one honest Verifier is $(1 - \bar{P}_\xi)^\Xi$. 
Consequently, the probability an error in \emph{any chunk in the block} remaining undetected is
\begin{align}\label{theorem:ExecutionFlow:CorrectnessProof:error-prob-exact}
	P_\textrm{error} = 1 - (1 - \bar{P}_\xi)^\Xi.
\end{align}
We assume that the system parameter $\eta$ is chosen such that  $\bar{P}_\xi \approx 0$ to ensure computational safety.
Hence, we can approximate eq.\ \eqref{theorem:ExecutionFlow:CorrectnessProof:error-prob-exact} by its first order Taylor series in $\bar{P}_\xi$.
\begin{align}\label{theorem:ExecutionFlow:CorrectnessProof:helper-expansion}
	(1 - \bar{P}_\xi)^\Xi \simeq 1 - \Xi \cdot  \bar{P}_\xi
\end{align}
Inserting equations \eqref{theorem:ExecutionFlow:CorrectnessProof:helper-expansion} and 
\eqref{theorem:ExecutionFlow:CorrectnessProof:numberChunksToBeChecked}	into \eqref{theorem:ExecutionFlow:CorrectnessProof:error-prob-exact} yields 
\begin{align}\label{theorem:ExecutionFlow:CorrectnessProof:block_error_probability_from_proof} 
	P_\textrm{error} \lesssim \Xi \cdot  \big(1 - \eta\big)^{\widetilde{N}},
\end{align}
which proves equation \eqref{theorem:ExecutionFlow:CorrectnessProof:block_error_probability} from the theorem. 
We have now shown that the probability of a faulty chunk in an \techterm{Execution Receipt} not being checked by an honest Verifier is bounded by \eqref{theorem:ExecutionFlow:CorrectnessProof:block_error_probability_from_proof}.
Furthermore, every honest Verifier will challenge any faulty chunk it is assigned to check by raising a \texttt{FaultyComputationChallenge} (see section \ref{sec:AttackMitigation:FaultyComputationResult} for details). 
Corollary \ref{corollary:blocksealing} guarantees that a block is only sealed if no correct \texttt{FaultyComputationChallenge}  was raised. 
Hence, the only way a block can be sealed with a faulty \texttt{ExecutionResult} is if the faulty chunks are not checked by honest Verifers. 
Consequently, eq.\ \eqref{theorem:ExecutionFlow:CorrectnessProof:block_error_probability_from_proof} also bounds the probability of a faulty \texttt{ExecutionResult} being sealed. 
\vspace{-7pt}
\begin{flushright}$\square$	\end{flushright}

\subsection{Computational Load on Verification Nodes\label{sec:ExecutionFlow:CompLoadOnVerifier}}

Using equation \eqref{theorem:ExecutionFlow:CorrectnessProof:block_error_probability}, we can compute the required fraction $\eta$ of chunks that 
each Verifier has to check to achieve a specific $P_\textrm{error}$.
For the mature system under full load, we expect there to be 1000 Verification Nodes and each block to contain up to $\Xi = 1000$ chunks. 
Furthermore, we make the conservative assumption that only $\frac{2}{3}$ of the Verification Nodes are honest, i.e.,\ $\widetilde{N}=667$. 

Let the probability for a malicious Execution Node to succeed with an attempt to introduce a compromised state into the blockchain be $P_\textrm{error} \leq 10^{-6}$.
To achieve this, every Verification Node would need to check $32$ chunks, i.e.,\ execute $\eta = 3.2\%$ of the work of an Execution Node. 
To lower the probability even further to $P_\textrm{error} \leq 10^{-9}$, Verifiers only need to  execute $\eta = 4.2\%$ of the work of an Execution Node. 

This shows that distributing and parallelizing the verification of computation results is efficient. 
Furthermore, note that checking the chunks can be trivially executed in parallel.

\clearpage
\section{Mitigation of Attack Vectors\label{sec:AttackVectors}}
In the following, we will discuss the most severe attacks on \Architecture. 
In this context, we would like to re-iterate that the staking balances are maintained by the Consensus Nodes as part of the network state 
(compare section \ref{sec:Computational_vs_Network-State}). Hence, Consensus Nodes can adjudicate and slash misbehaving nodes 
directly. The resulting updates to the network state are published in the blocks (field \texttt{slashingChallenges} in message \ref{msg:Block})
without needing to involve Execution Nodes.

\subsection{Adjudicating with Faulty Computation Results\label{sec:AttackMitigation:FaultyComputationResult}}

In section \ref{sec:ExecutionFlow:CorrectnessProof}, Theorem \ref{theorem:ExecutionFlow:CorrectnessProof}, 
we have shown that a faulty computation state will be challenged by a Verification Node with near certainty. 
Formally, a Verification Node submits a \techterm{Faulty Computation Challenge} (FCC), to the Consensus Nodes for adjudication.  
We start by introducing the necessary notation and then proceed 
with specifying  the details of an FCC and the protocol for processing them. 
\medskip

\begin{wrapfigure}{R}{0.51\textwidth}
	\begin{minipage}{0.51\textwidth}
		\vspace{-22pt}
		\begin{algorithm}[H]
			\renewcommand{\thealgorithm}{}
			\floatname{algorithm}{}
			\begin{algorithmic}[1]
				\State \textbf{message} FaultyComputationChallenge $\{$
				\State \quad \textbf{bytes}  executionReceiptHash;
				\State \quad \textbf{uint32} chunkIndex;
				\State \quad \textbf{ProofOfAssignment}  proofOfAssignment;
				\State \quad \textbf{StateCommitment}  stateCommitments;			
				\State \quad \textbf{Signature} verifierSig; 
				\State $\}$
			\end{algorithmic}
		\end{algorithm}
	\end{minipage}
	\captionof{messagedef}{
		Verification Nodes send this message  to Consensus Nodes to challenge a specific Execution Receipt (\texttt{executionReceiptHash}). 
		The \texttt{FaultyComputationChallenge} is specific to a computational output state for one of the chunks, where
		\texttt{chunkIndex} is a zero-based index into \texttt{ExecutionReceipt.executionResult.chunks} (compare Message \ref{msg:ExecutionReceipt}).
	}
	\label{msg:FaultyComputationChallenge}
	\vspace{-5pt}
\end{wrapfigure}

\begin{figure}[t!]
	\centering
	\vspace{-10pt}
	\includegraphics[width=0.95\textwidth, origin=c, trim=0 0 32 0, clip]{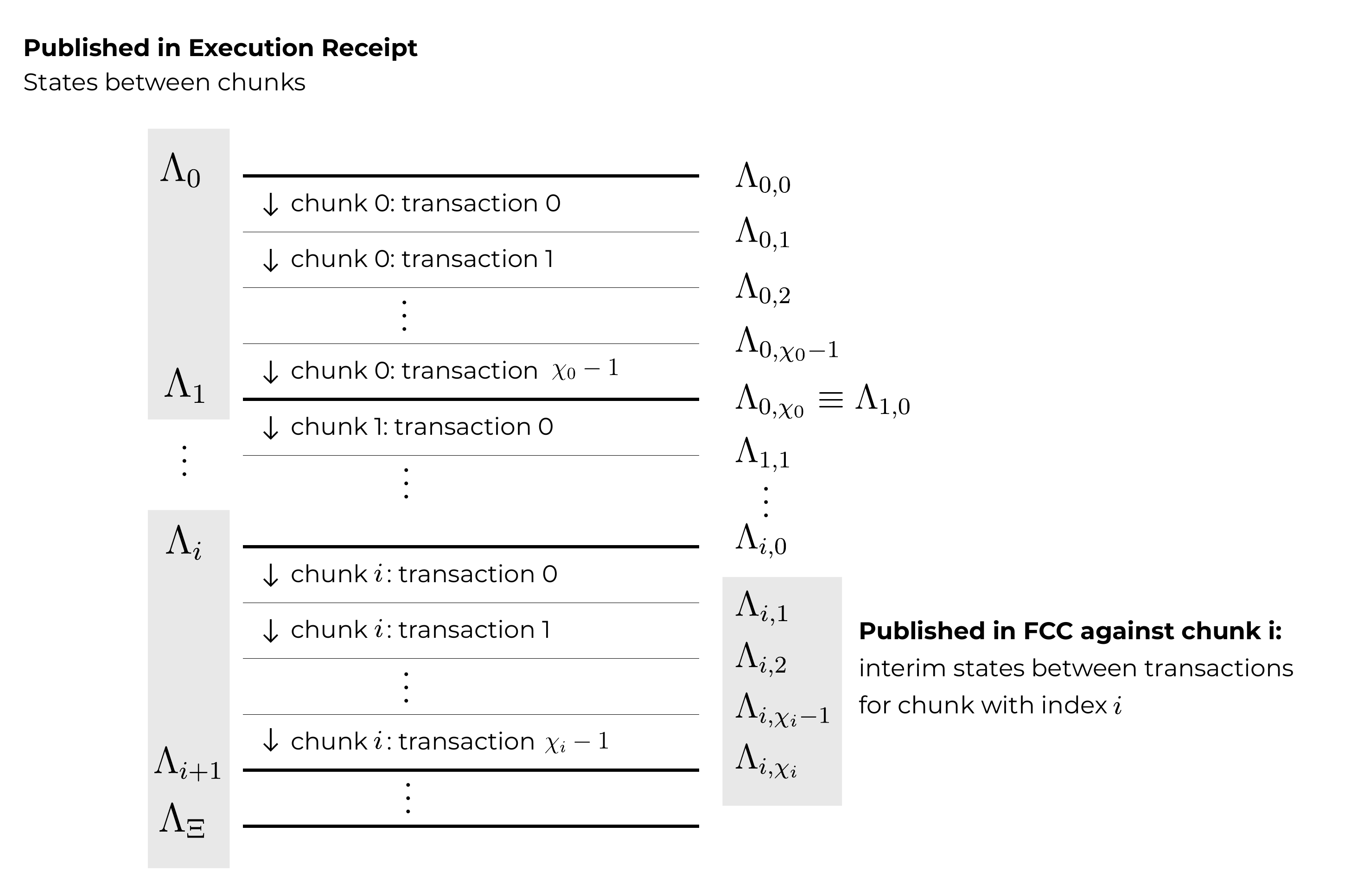}
	\vspace{-10pt}
	\caption{
		Illustration of nomenclature
		State commitments (e.g.,\ hashes) are represented as vertical lines and denoted by $\Lambda$. 
		The bold lines visualize the starting states $\Lambda_0, \Lambda_1, \ldots, \Lambda_{\Xi-1}$ for the chunks, 
		as well as the final state $\Lambda_{\Xi}$. 
		Furthermore, $\Lambda_{i,k}$ is the \emph{input} state for computing the transaction with index $k$ within the chunk. 
	}
	\label{fig:nomenclature:states_chunks}
\end{figure}

\noindent
\textit{Nomenclature} (illustrated in Figure \ref{fig:nomenclature:states_chunks}):  
For an Execution Receipt with $\Xi$ chunks, the field 

\noindent
\texttt{ExecutionReceipt.executionResult.chunks}
holds the \texttt{StateCommitment}s $[\Lambda_0, \Lambda_1, \ldots, \Lambda_\Xi]$. 
For $i \in \{0,1,\ldots, \Xi -1\}$, $\Lambda_i$  denotes the 
starting state for the computation of the chunk with index $i$.
$\Lambda_\Xi$ is the final state at the end of the block (after all
transactions are computed).

To denote the (interim) states  \emph{between individual transactions}, 
we extend the notation accordingly. Let the chunk at index $i$ contain $\chi_i$ transactions. 
For $k \in \{0,1,\ldots,\chi_i-1\}$, $\Lambda_{i,k}$ denotes the input state for computing the transaction with index $k$ within the chunk. 
Accordingly, $\Lambda_{i,\chi_i-1}$ is the state at the end of the chunk.

Note that $\Lambda_{i,\chi_i-1}$ simultaneously serves as the \emph{starting state} for the next chunk at index $i+1$.  
Hence, $\Lambda_{i,\chi_i}$ as well as   $\Lambda_{i+1}$ and $\Lambda_{i+1,0}$ refer to the same \texttt{StateCommitment}. 
The different nomenclatures are introduced for ease of notation only. 

\begin{definition}$~$\\
	A well-formed \textbf{Faulty Computation Challenge (FCC)}, specified in  Message \ref{msg:FaultyComputationChallenge}, 
	challenges one specific chunk at index \texttt{chunkIndex} of an Execution Receipt (referenced by \texttt{executionReceiptHash}). It provides the list
	\begin{align}
		\texttt{stateCommitments} = [\Lambda_{\texttt{chunkIndex}, 0}, \Lambda_{\texttt{chunkIndex}, 1}, \ldots, \Lambda_{\texttt{chunkIndex}, \chi_\texttt{chunkIndex}}].
	\end{align}  
\end{definition}

\newpage
\begin{definition}\label{def:AttackMitigation:Protocol:FaultyComputationResult}
\textit{Protocol for Adjudicating a Faulty Computation Result}\\
Let there be a Verifier Node $V$ that is checking the chunk at index $i$ and disagrees with the resulting state  $\Lambda_{i+1} \equiv \Lambda_{i, \chi_i}$. 
In the following, we denote the Verifier's \texttt{StateCommitment}s as  $\widetilde{\Lambda}$ and the Execution Node's with $\Lambda$.
\begin{enumerate}
	\item Verifier $V$ broadcasts a \texttt{FaultyComputationChallenge} to the Consensus Nodes with 
	\begin{itemize}
		\item \texttt{FaultyComputationChallenge.chunkIndex}  \textleftarrow \ $i$
		\item \texttt{FaultyComputationChallenge.stateCommitments}  \textleftarrow \ $[\widetilde{\Lambda}_{i, 1}, \ldots, \widetilde{\Lambda}_{i, \chi_i}]$
	\end{itemize}
	\item Consensus Nodes publish the FCC in their next finalized block (field \texttt{slashingChallenges} in message \ref{msg:Block})
	\item 
	The challenged Execution Node has a limited time to broadcast a \texttt{FaultyComputationResponse} 
	(Message \ref{msg:FaultyComputationResponse}) to the Consensus Nodes. 
	Time is measured using a \techterm{verifiable delay function} \cite{Boneh:2019:VDFs}.
	
	\begin{figure}[b!]
		\begin{center}
			\begin{minipage}{0.7\textwidth}
				\vspace{-5pt}
				\begin{algorithm}[H]
					\renewcommand{\thealgorithm}{}
					\floatname{algorithm}{}
					\begin{algorithmic}[1]
						\State \textbf{message} FaultyComputationResponse $\{$
						\State \quad \textbf{bytes}  FaultyComputationChallengeHash;
						\State \quad \textbf{StateCommitment}  stateCommitments;			
						\State \quad \textbf{Signature} executerSig; 
						\State $\}$
					\end{algorithmic}
				\end{algorithm}
				\vspace{-15pt}
			\end{minipage}
			\captionof{messagedef}{
				A challanged Execution Node broadcast a \texttt{FaultyComputationResponse} to the Consensus Nodes.
			}
			\label{msg:FaultyComputationResponse}
			\vspace{-10pt}
		\end{center}
	\end{figure}
	
	\begin{enumerate}
		\item Should the Execution Node not respond,  it is slashed. Consensus Nodes will include a corresponding notification in the next block 
		(field \texttt{networkStateUpdates}  in Message \ref{msg:Block}) that also includes the output of the VDF as proof of waiting. 
		In this case, adjudication ends with the Execution Node  being slashed.
		\item
		To prevent being slashed, the Execution Node must disclose all interim states in the chunk  by submitting a \texttt{FaultyComputationResponse} with
		\begin{itemize}
			\item \texttt{FaultyComputationResponse.stateCommitments}  \textleftarrow \ $[\Lambda_{i, 1}, \ldots, \Lambda_{i, \chi_i}]$
		\end{itemize}		
		to the Consensus Nodes.
		 In case the Execution Node sends a \texttt{FaultyComputationResponse}, the protocol continues with step \ref{AttackMitigation:MissingCollection:FaultyComputationAdjudication_0} below.
	\end{enumerate}	
	\item \label{AttackMitigation:MissingCollection:FaultyComputationAdjudication_0}
	Consensus Nodes now compare the \texttt{stateCommitments} from both parties element-wise and find the \emph{first} mismatch. 
	Let the first mismatch occur at index $\ell$, i.e.
	\begin{align}
			& \Lambda_{i, \ell} \neq \widetilde{\Lambda}_{i, \ell} \\
			\textrm{for all } l \in [0,1,\ldots, \ell-1]: \quad & \Lambda_{i, l} =\widetilde{\Lambda}_{i, l}.
	\end{align}
	Essentially, both parties agree that, starting from state $\Lambda_{i} \equiv \Lambda_{i,0}$, the computation should lead to 
	 $\Lambda_{i, \ell-1}$ as the input state for the next transaction.
	However, they disagree on the resulting state \emph{after} computing this next transaction.
	\item
	Consensus Nodes request state $\Lambda_{i, \ell-1}$ from either party.
	Furthermore, by resolving the texts of collections in the block, Consensus Nodes obtain the transaction with index $\ell$ in the chunk, 
	 whose output is disputed.
	\item 
	Consensus Nodes use $\Lambda_{i, \ell-1}$ as input state for computing transaction with index $\ell$ in the chunk. 
	Consensus Nodes now compare their locally computed output state with $\Lambda_{i, \ell}$ and $\widetilde{\Lambda}_{i, \ell}$.
	\item 
	Any party who published a computation result that does not match the values computed by the Consensus Nodes is slashed. 
	Consensus Nodes will include a corresponding notification in the next block 
	(field \texttt{networkStateUpdates}  in message \ref{msg:Block}).
\end{enumerate}
\end{definition}

\noindent
Informally, Definition \ref{def:AttackMitigation:Protocol:FaultyComputationResult} describes a protocol by which 
a Verifier Node can appeal to the committee of Consensus Nodes to re-compute a specific transaction whose output the Verifier does not agree with. 
To avoid being slashed, the challenged Execution node must provide all information that is required for the Consensus Nodes to re-compute the transaction in question.
Nevertheless, there is no leeway for the Execution Node to provide wrong information as honest Consensus Nodes will verify the correctness 
based on previously published hash commitments:
\begin{itemize}
	\item Consensus Nodes request transaction texts of collections. The collection hashes are stated in blocks, which allow  verification of the collection texts. 
	\item Consensus Nodes request the last interim state $\Lambda_{i, \ell-1}$ in the chunk that both parties agree on. 
	           A hash of this state was published by both the Verifier and the challenged Execution Node. This allows the Consensus Nodes verify state variables (e.g.,\ via Merkle proofs).
\end{itemize}
Furthermore, the described protocol is executed by \emph{all} honest Consensus Nodes. The resulting slashing is included in a block and hence secured by BFT consensus.
Assuming a super-majority of honest Consensus Nodes, it is guaranteed that the misbehaving node is slashed. 
The following Corollary \ref{corollary:AttackMitigation:FaultyComputationResult} formalizes this insight. 

\newpage 
\begin{corollary}\label{corollary:AttackMitigation:FaultyComputationResult}
	$~$\\
	Given a system with
	\begin{itemize}
		\item more than $\frac{2}{3}$ of the Consensus Nodes' stake is controlled by honest actors;
		\item and partially synchronous network conditions with message traversal time bounded by $\Delta_t$.
	\end{itemize}
	The following holds.
	\begin{itemize}
		\item 	
		If an honest Verifier  node is assigned to verify a chunk that has a faulty computation result,
		the Execution Node who issues the  corresponding Execution Receipt will be slashed. 
		\item 
		If a dishonest Verifier Node challenges a correct computation result, the Verifier will be slashed. 
	\end{itemize}
	
\end{corollary}

\subsection{Resolving a Missing Collection\label{sec:AttackMitigation:MissingCollection}}

As message  \ref{msg:Block}  and \ref{msg:Collection} show, a block references its collections by hash but does not contain the individual transactions texts. 
The transactions texts are stored by the Collector Node cluster which build the collection and is only required when Execution Nodes want to compute the blocks' transactions.
Hence, a cluster of Collector nodes could withhold the transaction texts for a guaranteed collection. 
While block production is not impacted by this attack, block execution halts without access to the needed transaction texts. 

When an Execution Node is unable to resolve a guaranteed collection, it issues a \techterm{Missing Collection Challenge} (MCC). 
An MCC is a request to slash the cluster of Collector Nodes (Message \ref{msg:Collection}, line \ref{msg:Collection:aggregatedCollectorSigs}) who have guaranteed the missing collection. 
MCCs are directly submitted to Consensus Nodes.

\begin{definition}\textit{Protocol for Resolving a Missing Collection}
	\begin{enumerate}
		\item 
		An Execution Node determines that the transaction texts for a \texttt{GuaranteedCollection} from the current block are not available as expected. 
		The protocol does not dictate how this determination is reached, but the obvious implementation is assumed (ask the Guarantors, wait for a response, ask other Execution Nodes, wait for a response).
		\item 
		The Execution Node broadcasts an MCC to all Collector  and Consensus Nodes. 
		The Consensus Nodes record the challenge in the next block, but do not otherwise adjudicate the challenge at this stage.
		\item \label{AttackMitigation:MissingCollection:HonestGuarantorProtocol_0}
		Any honest Collector Node, which is \emph{not} a member of the challenged cluster, 
		sends a request to $\kappa$ randomly selected Guarantors to provide the missing Collection. 
		If the request is answered, the requesting Collector Node forwards the result to the Execution Nodes.
		\item 
		If the Guarantor does not respond within a reasonable time period $\Delta_t$, 
		the requesting Collector Node will sign a \techterm{Missing Collection Attestation} (MCA), including the hash of the collection in question. 
		Time is measured using a \techterm{verifiable delay function} \cite{Boneh:2019:VDFs} and the MCA contains the VDF's output as proof of waiting.
		 % \note{This might be formalized further: explicitly state algorithm and mechanism to prevent copying of VRF output.}
		The MCA is broadcast to all Consensus and Execution Nodes.
		\item \label{AttackMitigation:MissingCollection:HonestGuarantorProtocol_1}
		An honest challenged Guarantor will respond with the complete Collection text to any such requests. 
		\item 
		If the Execution Nodes receive the collection, they process the block as normal. 
		Otherwise, they wait for more than $\frac{2}{3}$ of the Collector Nodes to provide MCAs. 
		\item 
		Appropriate MCAs must be included in all Execution Receipts that skip one or more Collections from the block.
		\item 
		Every MCC will result in a small slashing penalty for each Execution Node and each challenged Guarantor. 
		Even if the MCC is resolved by finding the Collection, each of these actors must pay the fine, including the Execution Nodes that did not initiate the MCC. 
		This is designed to prevent the following edge cases: 
		\begin{itemize}
			\item 
			Lazy Guarantors that only respond when challenged: without a fine for challenged Guarantors, 
			even in the case where the collection is resolved, there is no incentive for Guarantors to respond without being challenged. 
			\item 
			Spurious MCCs coming from Byzantine Execution Nodes: without a fine for Execution Nodes, there is a zero-cost  to generating system load through unjustified MCCs.
			\item Don't punish the messenger: all Execution Nodes must be fined equally so that there is no disincentive to be the first Execution Node to report a problem.
		\end{itemize}
		\item 
		If an Execution Receipt containing an MCC is sealed, ALL guarantors for the missing Collection are subject to a large slashing penalty 
		(equal to the minimum staking requirement for running a Collector Node).
	\end{enumerate}
\end{definition}

\subsubsection*{Discussion}
\begin{itemize}
	\item 
	The slashing penalty for a resolved MCC should be small enough that it doesn’t automatically eject the Collector Nodes from the network 
	(by dropping them below the minimum staking threshold), but must be significant enough to account for the fact that resolving an MCC is very expensive.
	\item 
	Resolving an MCC is very expensive. Each Collector Node will request the Collection from one Guarantor, 
	so each Guarantor will have to respond to many requests or risk being slashed. Each Execution Node will be flooded with copies of the Collection if it is available. 
	We are operating under the theory that if MCCs have a very high probability of being resolved correctly (Lemma \ref{theorem:AttackMitigation:MissingCollection}), 
	spurious MCCs should be very rare specifically because of the described penalties. 
	\item 
	If most Execution Nodes are Byzantine and raise spurious MCCs, 
	but at least one Execution Node is honest and generates complete Execution Receipts, a correct Execution Receipt will be sealed 
	(assuming an honest super-majority of Collector Nodes and Consensus Nodes). 
	Furthermore, the Execution Nodes who raised the spurious MCC will be slashed. 
	\item 
	If most Guarantors of a particular Collection are Byzantine, and refuse to provide Collection data, 
	but at least one Guarantor is honest, the Collection will be provided to an honest Execution Node and executed properly.
	\item 
	A cabal of 100\% of Execution Nodes acting maliciously can halt the network by not executing new blocks.
	Nevertheless, no faulty state can be introduced into the network by such a denial of service attack.
	\item 
	In order for an attacker to obtain the ability to introduce an error into the computation state (with non-negligible probability),
	the attacker would need to control a Byzantine fraction of 100\% of Execution Nodes 
	\emph{and} more than $\frac{2}{3}$ of Verifier Nodes.
\end{itemize}

\begin{theorem}\label{theorem:AttackMitigation:MissingCollection}
	\textit{\textbf{Liveness of Collection Text Resolution}}\\
	Given a system with
	\begin{itemize}
		\item more than $\nicefrac{2}{3}$ of the Consensus Nodes' stake is controlled by honest actors;
		\item more than $\nicefrac{2}{3}$ of the Collector Nodes' stake is controlled by honest actors;
		\item and partially synchronous network conditions with message traversal time bounded by $\Delta_t$.
	\end{itemize}
    The system can be configured such that any guaranteed collection is available with probability close to $1$.
\end{theorem}

\noindent%
\textbf{Proof of Theorem  \ref{theorem:AttackMitigation:MissingCollection}}\\
For this proof, we assume that Byzantine nodes collaborate to the maximum extent possible to prevent a collection from being resolved. 
Unless there are protocol-level guarantees, we consider the most favorable conditions for the Byzantine nodes.
Let us assume that the Byzantine nodes successfully prevented a collection from being resolved, 
i.e.,\ more than $\nicefrac{2}{3}$ of the collectors issued an MCA.
Let
\begin{itemize}
	\item $N$ the total number of collector nodes, $\widetilde{N}$ the number of honest collectors, and $\bar{N}$ the number of Byzantine collectors;
	\item $n_\textrm{cluster}$ the size of the collector cluster that produced the missing collection,
	           $\tilde{n}_\textrm{cluster}$ the number of honest collectors in the cluster, and $\bar{n}_\textrm{cluster}$ the number of Byzantine collectors in the cluster;
	\item $n_\textrm{g}$ the number of guarantors of the missing collection,
	           $\tilde{n}_\textrm{g}$  the number of honest guarantors, and $\bar{n}_\textrm{g}$ the number of Byzantine guarantors.
\end{itemize}
We consider a system configuration with $N = \widetilde{N}  + \bar{N} = 1000$ collector nodes where $\widetilde{N} = 2 \floor{N / 3} + 1 $.
In \Architecture, the clusters are created by randomly partitioning the collectors into clusters of size $n_\textrm{cluster}$ 
via Fisher-Yates shuffling. Hence, the probability of drawing a cluster with $\bar{n}_\textrm{cluster}$ Byzantine actors is given by the hypergeometric distribution
\begin{align}
\mathcal{P}_{_{n_\textrm{cluster},N,\bar{N}}}(\bar{n}_\textrm{cluster} ) = \frac{{\bar{N} \choose \bar{n}_\textrm{cluster} } {N-\bar{N} \choose n_\textrm{cluster}-\bar{n}_\textrm{cluster} }}{{N \choose n_\textrm{cluster}}} \, .
\end{align}
For a collection, at least $n_\textrm{g} = 2 \floor{n_\textrm{cluster} / 3} + 1 = \bar{n}_\textrm{g}  + \tilde{n}_\textrm{g}$ guarantors are required. 
The number of Byzantine guarantors $\bar{n}_\textrm{g}$ could take any value in $0,1,\ldots, \bar{n}_\textrm{cluster}$.
There could be more Byzantine nodes in the cluster than required to guarantee the collection, i.e., $n_\textrm{g} < \bar{n}_\textrm{cluster}$.
In this case, we assume that only the minimally required number of  $n_\textrm{g}$ Byzantine nodes would guarantee the collection  to minimize slashing.
\begin{align}
\bar{n}_\textrm{g} \in \Big\{0, 1, \ldots, \min\big(n_\textrm{g}, \, \bar{n}_\textrm{cluster} \big) \Big\}\,.
\end{align}
As each honest guarantor increases the probability of a collection being successfully retrieved,
we assume that the Byzantine nodes only involve the absolute minimum number of honest nodes to get the collection guaranteed:
\begin{align}
	\tilde{n}_\textrm{g} & = n_\textrm{g}  - \bar{n}_\textrm{g} = \, 2 \floor{n_\textrm{cluster} / 3} + 1 - \bar{n}_\textrm{g}
\end{align}

\begin{figure}[t!]
	\centering
	\includegraphics[trim=0 10 0 24, clip, width=0.7\textwidth]{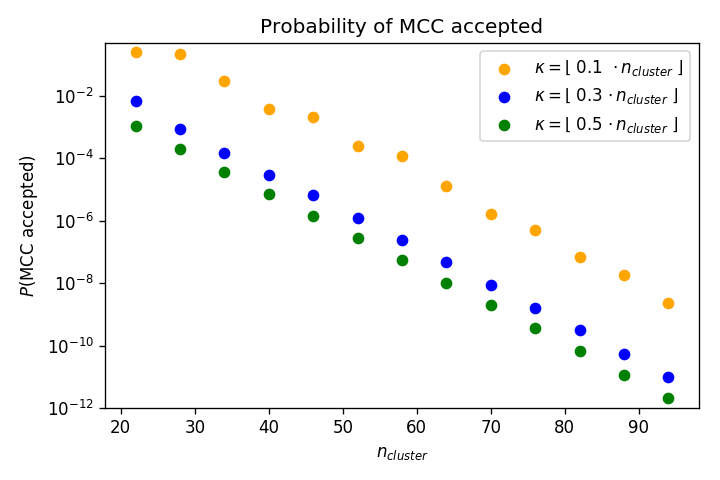}
	\caption{
		\textbf{Worst-case probability that a collection cannot be resolved}. 
		The graph shows numerical values for $P(\textnormal{MCC accepted})$, equation \eqref{theorem:AttackMitigation:MissingCollection:Prob_MCC_Accepted}, 
		for $N = 1000$, $\widetilde{N} = 667$ and $\bar{N} = 333$.
	}
	\label{fig:MCC_accepted_prob_scaling}
\end{figure}

When an honest collector that is \emph{not} a guarantor receives a MCC, it selects $\kappa$ guarantors and requests the collection from them.
We assume that only honest guarantors would answer such a request.
The probability for a correct node to receive no answer when inquiring about the missing collection, i.e., to issue in MCA, is
\begin{align}
\mathcal{P}_{_{\kappa,n_\textrm{g},\tilde{n}_\textrm{g}}}(0) = \frac{{n_\textrm{g}-\tilde{n}_\textrm{g}\choose \kappa} }{{n_\textrm{g} \choose \kappa}} \, .
\end{align}
Furthermore, every Byzantine node that is \emph{not} a guarantor of the collection would issue an MCA to increase the chances that the \techterm{Missing Collection Challenge}  is accepted.
Hence, there are $(\bar{N} - n_\textrm{g})$ MCAs from Byzantine nodes. 
For a collection to be considered missing, the protocol require $\frac{2}{3} N$ collectors to send MCAs.
Consequently, the minimally required number of MCAs from honest nodes is
\begin{align}
  r = \widetilde{N} - (\bar{N} - \bar{n}_\textrm{g}).
\end{align}
As honest nodes independently contact guarantors,
each honest node conducts a Bernoulli trial and issues a MCAs with probability $\mathcal{P}_{_{\kappa,n_\textrm{g},\tilde{n}_\textrm{g}}}(0)$.
Consequently, the probability that $r$ honest nodes  issue  MCAs follows a Binomial distribution
\begin{align}
\mathcal{B}(r) = {\widetilde{N}  \choose r}  \big[ \mathcal{P}_{_{\kappa,n_\textrm{g},\tilde{n}_\textrm{g}}}(0) \big]^r \big[ 1 - \mathcal{P}_{_{\kappa,n_\textrm{g},\tilde{n}_\textrm{g}}}(0) \big]^{\widetilde{N} - r}\,.
\end{align}
Given the number $\bar{n}_\textrm{cluster}$ of byzantine actors in the cluster,  
the worst-case probability that the MCC is accepted by the system is
\begin{align}
P(\textnormal{MCC accepted}\, | \, \bar{n}_\textrm{cluster})  = 
\underset{\bar{n}_\textrm{g} \in  \atop  \{0, 1, \ldots, \min(n_\textrm{g}, \, \bar{n}_\textrm{cluster}) \}}{\operatorname{max}} \quad 
\sum_{r = \atop \widetilde{N} - \bar{N} + \bar{n}_\textrm{g}}^{\widetilde{N}} \mathcal{B}(r)\,.
\end{align}
The overall probability of an MCC being accepted is, therefore,
\begin{align}\label{theorem:AttackMitigation:MissingCollection:Prob_MCC_Accepted}
P(\textnormal{MCC accepted})   =  \sum_{\bar{n}_\textrm{cluster}  \in \atop \{ 0,\ldots, n_\textrm{cluster}\}} P(\textnormal{MCC accepted}\, | \, \bar{n}_\textrm{cluster})   \cdot \mathcal{P}_{_{n_\textrm{cluster},N,\bar{N}}}(\bar{n}_\textrm{cluster} )
\end{align}
Figure \ref{fig:MCC_accepted_prob_scaling} illustrates the worst-case probability of a chunk missing, i.e., that a MCC 
shows the probabilities according to equation \eqref{theorem:AttackMitigation:MissingCollection:Prob_MCC_Accepted}.

\vspace{-15pt}
\begin{flushright}$\square$	\end{flushright}
	
\subsection{Placing errors in chunks checked by colluding Verifier Nodes\label{sec:attack-vectors:ColludingENandVNs}}

If an Execution Node and several Verifier Nodes are colluding, they have the ability to secretly determine which chunks would be checked 
by the colluding Verifiers before even publishing an Execution Receipt. 
However, the self-assignment scheme defined in section \ref{sec:ExecutionFlow:Verification} is independent for each Verifier 
and in addition non-predictable for anyone without the Verifier's private key.
Therefore, honest Verifiers will still check each chunk with uniform probability,  independently of the colluding Verifiers.
Consequently, if a malicious Execution Node wants to introduce a computation error, there is no advantage in placing the 
error in chunks that are checked by colluding Verifiers. This insight is formalized as Corollary \ref{corollary:AttackMitigation:ColludingENandVNs}.

\begin{corollary}\label{corollary:AttackMitigation:ColludingENandVNs}
	$~$\\
	Given a system with partially synchronous network conditions with message traversal time bounded by $\Delta_t$.
	Let there be a malicious Execution Node  that tries to  introduce a computation error into one of the chunks of a block. 
	The success probability cannot be increased by the chunk selection of Byzantine Verifiers.
\end{corollary}

\clearpage
\addcontentsline{toc}{section}{Acknowledgments}
\section*{Acknowledgments}

We thank Dan Boneh for many insightful discussions, 
and 
Alex Bulkin, 
Karim Helmy, 
Chris Dixon,
Jesse Walden,
Ali Yahya,
Ash Egan,
Joey Krug,
Arianna Simpson,
as well as  Lydia Hentschel
for comments on earlier drafts. 

\addcontentsline{toc}{section}{References}
\footnotesize
\bibliography{post-computation}{}
\bibliographystyle{unsrt} 

\end{document}